\documentclass[aps,prd,twocolumn,showpacs,groupedaddress,nofootinbib
]
{revtex4-1}

\expandafter\let\csname equation*\endcsname\relax 
\expandafter\let\csname endequation*\endcsname\relax

\usepackage{times}
\usepackage{amsfonts}
\usepackage{graphicx}
\usepackage{amsmath}
\usepackage{amssymb}
\usepackage{color}
\usepackage{subfigure}
\usepackage{natbib}
\usepackage{tensor}
\usepackage{accents}
\usepackage[colorlinks=true, citecolor=blue]{hyperref}
\usepackage{xspace}

\newcommand{\lie}{\pounds}
\newcommand{\eom}{\mathcal{E}}
\newcommand{\fom}{\mathcal{F}}

\newcommand{\constr}{\mathcal{C}}

\def\beq{\begin{equation}}
\def\eeq{\end{equation}}
\def\bea{\begin{eqnarray}}
\def\eea{\end{eqnarray}}
\def\ben{\begin{enumerate}}
\def\een{\end{enumerate}}

\def\a{\alpha}

\def\d{\delta}
\def\e{\epsilon}

\def\k{\kappa}
\def\l{\lambda}

\def\t{\tau}

\def\half{{\textstyle{\frac{1}{2}}}}
\def\third{{\textstyle{\frac{1}{3}}}}

\def\Jt{\tilde{J}}
\def\et{\tilde{\epsilon}}

\def\vphi{\varphi}

\def\Horava{Ho\v{r}ava\xspace}

\begin{document}

\title{Variations on an aethereal theme }

\author{Ted Jacobson}
\email{jacobson@umd.edu}
\affiliation{Maryland Center for Fundamental Physics, 
University of Maryland, College Park, MD 20742}
\affiliation{Perimeter Institute for Theoretical Physics, 31 Caroline Street North, Waterloo,
ON N2L 2Y5, Canada}

\author{Antony J. Speranza}
\email{asperanz@umd.edu}
\affiliation{Maryland Center for Fundamental Physics, 
University of Maryland, College Park, MD 20742}
\affiliation{Perimeter Institute for Theoretical Physics, 31 Caroline Street North, Waterloo,
ON N2L 2Y5, Canada}

\date{June 16, 2015}

\begin{abstract}

We consider a class of Lorentz-violating theories of gravity involving a timelike unit vector field
(the aether) coupled to a metric, two examples being Einstein-aether theory and \Horava gravity.
The action always includes the Ricci scalar of the metric and the invariants
quadratic in covariant derivatives of the aether, but the theories differ in how the aether is constructed
from other fields, and whether those fields are varied in the action. Fields that are not varied
define background structures breaking diffeomorphsim invariance, including threadings, folations, 
and clocks, which generally produce novel degrees of freedom arising from the violation of what 
would otherwise be initial value constraints. The principal aims of this paper are to survey the
nature of the theories that arise, and to understand the consequences of breaking 
diffeomorphism invariance in this setting. 
In a companion paper [arXiv:1504.03305], 
we address some of the phenomenology of the ``ponderable 
aether'' case in which the presence of a background clock endows the aether with a variable 
internal energy density that behaves in some respects like dark matter.

\end{abstract}

\maketitle

\section{Introduction and summary}

Longstanding puzzles of cosmology and quantum gravity 
have led some to question the fundamental assumption of general
relativity, that the spacetime manifold has no structure other
than that determined by the metric.
In particular, the cosmological constant problem, dark energy,
dark matter, the trans-Planckian puzzle, the need for 
a UV completion of general relativity, the problem of time
and the interpretation of quantum cosmology have motivated
exploration of modified gravity theories with vacuum structure
violating local Lorentz boost symmetry.  
If exact rotational symmetry
is preserved, a Lorentz violating vacuum structure selects a preferred
timelike direction at each spacetime point. The 
integral curves of this field of directions 
may be thought of as the flow 
of an ``aether fluid."
 The 4-velocity $u^a$ of the aether is
the unit timelike vector field  tangent to this 
flow. 

In constructing a theory with such an aether, one must decide whether 
the aether is to be treated as dynamical, i.e.\ varied
in the action principle, or instead as background. If the aether is
dynamical, then one must further specify how it is constructed in terms of the fields that are
varied in the action. Actually it turns out that the distinction between varied and not varied
fields is not so clear cut: the equations of motion for scalar fields are often  
a consequence of the equations of motion of the other fields. Such scalars, and the structures they
define, can therefore be regarded as ``background" structure, even though they might also be varied 
in the action. What is important for the physics, however, is not how we refer to them, but how 
these choices affect the degrees of freedom and behavior of the theories. The purpose of this
paper is to examine this question for a variety of related aether theories.

It is natural to assume that
the aether-metric dynamics is governed by an (effective) action  
involving the metric, its curvature, the aether, and covariant derivatives
of the aether and the curvature. 
Before beginning with the detailed analysis, we would like to 
point out that the theory would generally be dynamically overconstrained,
i.e.\ ``inconsistent," if derivatives of the aether were not included in the action. 
Suppose for example the action $S[g_{ab},\phi,u^a]$ is a scalar constructed from 
the metric $g_{ab}$, a scalar field $\phi$, and a unit vector field $u^a$.
Suppose further that the aether enters the Lagrangian density only via the  
coupling
$\half \sqrt{-g}(u^a\phi_{,a})^2$. The variations of $u^a$ must be orthogonal to $u^a$
to preserve the unit condition, and these impose the equation of motion 
$(\d S/\d u^a)(\d^a_b -u^a u_b)= \sqrt{-g}(u^m\phi_{,m})[\phi_{,b} -(u^n\phi_{,n})u_b]=0$.
This extremely restrictive condition requires that either $\phi$ is constant along the flow
lines of $u^a$, or the flow of $u^a$ is hypersurface orthogonal and $\phi$  
is constant on the orthogonal hypersurfaces. This eliminates virtually all of the solutions
to the scalar equation of motion. Moreover, even if we choose to not impose the 
aether equation of motion, the scalar field is still overconstrained since, as shown below,
the other equations of motion imply that the Lie derivative $\lie_u[(\d S/\d u^a)(\d^a_b -u^a u_b)]$
vanishes. Although a weaker condition, this still eliminates almost all scalar field solutions.
The situation is quite different, however, if the action includes terms quadratic in 
aether derivatives. Then $\d S/\d u^a$ includes second derivatives of the aether,
so instead of overconstraining the scalar field the extra conditions can be propagation 
equations for the aether.

At lowest order in a derivative expansion, the most general 
action for the metric and aether
is given (up to the integral of a total divergence) 
by\footnote{We use the metric signature $({+}{-}{-}{-})$. Abstract indices are denoted 
by Latin letters, spacetime coordinate indices by Greek letters, and
comma and semicolon before an index denote partial and covariant derivative respectively. 
Quantities with density weight 1 are written in caligraphic font, $\eom,\fom,\constr$, 
or carry a tilde, unless they involve the metric determinant 
or are written explicitly as a variational derivative.}
\begin{align}
S&[g_{ab},u^a(\zeta, g_{ab})] =\nonumber \\ 
&\frac{-1}{16\pi G_0} \int d^4x\sqrt{-g} (R+\frac{c_\theta}{3}\theta^2+c_\sigma\sigma^2+c_\omega \omega^2+c_a a^2), \label{eqn:action}
\end{align}
where $\zeta$ denotes collectively independently varied fields 
used in the construction of the aether $u^a(\zeta, g_{ab})$. 
The terms in the integrand are the Ricci scalar $R$ of the metric, 
and the expansion $\theta = \nabla_a u^a$, shear 
$\sigma_{ab} = \nabla_{(a}u_{b)}+\third (\nabla_c u^c)h_{ab}- u_{(a}a_{b)}$, 
twist $\omega_{ab} = \nabla_{[a}u_{b]}-u_{[a}a_{b]}$ 
and acceleration $a_b = u^a\nabla_au_b$ of the aether flow
(here $h_{ab}=u_au_b - g_{ab}$ is the spatial metric).
In this paper we examine how variations on the 
construction of $u^a(\zeta,g_{ab})$
affect the resulting theory.

An important distinction is
whether or not the aether vector is necessarily hypersurface-orthogonal, 
and hence non-twisting. 
The minimal structure required to determine a twist-free aether is
a \emph{foliation} by spacelike hypersurfaces, 
whereas the minimal structure required to determine a twisting aether is a
timelike congruence of curves, i.e.\ a \emph{threading} \cite{Boersma1995, Boersma1995a}. 
A well-known example of a foliation theory results if 
$u^a$ is constructed from a scalar field $T$ as
\beq\label{u(T)}
u^a(T,g_{ab}) = {g^{ab}T_{,b}}/{|dT|},
\eeq
where 
$|dT| = (g^{ab} T_{,a}T_{,b})^{1/2}$ is the norm of the gradient.
By construction, the aether \eqref{u(T)} is orthogonal to the constant
$T$ surfaces.
This is  khronometric
theory \cite{Blas2009, Blas2010, Jacobson2010a}, a.k.a.\ the infrared limit of the nonprojectable version  of \Horava{}-Lifshitz gravity \cite{Horava2009}. 
Among threading theories, a well-known example 
results
if $u^a$ is constructed from an independent vector field $A^a$ as
\beq\label{u(A)}
u^a(A^m,g_{mn}) = {A^a}/{|A|},
\eeq
where $|A|=(g_{mn}A^mA^n)^{1/2}$ is the norm of $A^a$.
This is Einstein-aether theory \cite{Jacobson2001, Jacobson2008}, written in the form given 
in ref.\ \cite{Jacobson2013}. In \eqref{u(A)} the unit constraint on $u^a$
holds by construction, rather than being imposed via a Lagrange multiplier term
as is more commonly done.
 
In the constructions we study,  
the foliation or threading is in most cases described covariantly using scalar fields which   
enter the construction of $u^a$ through their gradients. 
Since the 
action \eqref{eqn:action} involves first
derivatives of $u^a$,  it involves \emph{second} 
derivatives of the scalars,   
raising the concern
that the resulting theory might suffer from
an Ostrogradski instability \cite{Woodard2015}.
However, 
diffeomorphism invariance 
implies that the Lagrangian is degenerate, so that there is 
a possibility that the instability is absent.  
In fact, as we explain in section \ref{sec:prelims},
for all theories we consider, 
the scalar field equations are redundant with
the other field equations. Therefore the scalars need not be varied
in the action principle,
so we may fix their values by a choice of coordinates
at the level of the action. 
The gauge-fixed action is no longer invariant under the full group of 
diffeomorphisms, yet it gives a theory that is  equivalent to 
the one defined by the original diffeomorphism invariant 
action.\footnote{It is notoriously challenging to define what it means for a \emph{theory itself}---as opposed to its formulation---to be diffeomorphsim invariant \cite{BrianPitts2006, Pooley2015}.} 
There exists a coordinate gauge choice for which
the scalars' gradients have constant components,
and in such a gauge
the action is only first order in derivatives of the remaining dynamical variables.  
(For example, in the khronometric 
theory with one dynamical scalar $T$,  $u^a$ in \eqref{u(T)} is first order in derivatives, 
but with the gauge choice $x^0=T$ its components become  
$u^\alpha = g^{\alpha0}/\sqrt{g^{00}}$, which contain no derivatives.)
We conclude from this that there is no Ostrogradski instability in any gauge.

In
Einstein-aether theory, the 
threading is determined by a ``line field," i.e.\ a vector field modulo local scaling.
We call this the \textit{dynamical aether theory}, since it arises from a dynamical
field $A^a$ that appears with  only first derivatives in the action.
In subsection \ref{sec:bgthreads}, we consider a related theory where
the threading is determined by three
scalar fields, that can be fixed as background structures
as explained above.
We show that this \textit{fixed threading theory} is equivalent to Einstein-aether theory 
except that it admits violation of the 
spatial  
initial value constraints. 
The constraint violation is characterized by a spatial covector density that is 
preserved along the aether flow, and does not affect the energy-momentum tensor.

Subsection \ref{sec:clockthreads} considers a different theory, in which
the aether threading is 
determined by a line field as in Einstein-aether theory, and there is an additional 
scalar field which determines a preferred clock constrained to measure 
proper time along the threads.  
This \emph{fixed clock  theory} is equivalent to Einstein-aether theory 
except that it admits violation of 
a single
initial value constraint per spatial point. The constraint violation is 
characterized by a scalar density that is 
preserved along the aether flow, and appears in the energy-momentum tensor
like a rest mass density of the aether fluid. We call this a 
\emph{ponderable aether}, invoking the 19th century adjective that was used to 
distinguish ordinary matter from aether. Finally in subsection \ref{sec:bgaether}
we consider the \emph{fixed aether theory}, containing both a fixed 
threading and a fixed clock.  The four scalars that describe the background 
structure completely determine the aether vector, and hence this theory is equivalent
to Einstein-aether theory with the vector $u^a$  taken to be nondynamical. 

In section \ref{sec:foliation}, we consider foliation-type theories. After 
reviewing the khronometric theory, we consider in section \ref{sec:clockfoliation}
introducing an independent clock field.  As before this leads to a violated constraint 
as well as an additional component in the stress tensor.  Finally, we show
that when the clock field is constrained to coincide with the preferred foliation, 
the resulting theory is projectable \Horava gravity, and we again find a violated 
constraint. This constraint violation was already studied in 
ref.\ \cite{Mukohyama2009}, which referred to it as ``dark matter as an integration constant."
Unlike before, however, the violation in this case is not preserved under flows of $u^a$
for a generic form of the action \eqref{eqn:action}.

\section{Variational Preliminaries} \label{sec:prelims}
We begin by establishing the notation and several key results that are used throughout 
this paper.  
We define the following tensor densities, resulting from variations 
of the Einstein-aether action \eqref{eqn:action},
\begin{align}\label{EabFab}
\eom_{ab}&\equiv  \left. \frac{\delta S}{\delta g^{ab}} \right|_{u^c}, \quad
\fom_{ab}\equiv\left.\frac{\delta S}{\delta g^{ab}} \right|_{u_c},\\
\eom_c&\equiv \left.\frac{\delta S}{\delta u^c}\right|_{g^{ab}} = g_{cd}\left.\frac{\delta S}{\delta u_d}\right|_{g^{ab}}.\label{Ea}
\end{align}
The object to the right of the vertical line in these expressions indicates which tensors
are held fixed when computing the variation, i.e.\ $\eom_{ab}$ and $\fom_{ab}$ differ
in that contravariant $u^c$ is held fixed in the former, while covariant $u_c$ is fixed in
the latter.  They are related by 
\begin{equation}\label{eqn:EFreln}
\fom_{ab} = \eom_{ab}+\eom_{(a}u_{b)}.
\end{equation}

As discussed in the introduction, 
we are interested in cases where $u^a$ is constructed from $g_{ab}$ and other fields 
collectively denoted $\zeta$.  The Einstein equation results from varying the action holding
$\zeta$ fixed, and hence receives a contribution from the explicit metrics appearing in 
the action, as well as from the metric dependence of $u^a$.  In all
cases considered, $u^a$ depends algebraically on the metric, and the Einstein equation
takes the form
\begin{equation}\label{Einsteinu^c}
\eom_{ab} + \eom_c \frac{\delta u^c}{\delta g^{ab}} = 0.
\end{equation}
For the foliation-type theories of section \ref{sec:foliation}, it  
is more convenient to work with 
covariant $u_c$, in which case the expression for the Einstein equation is
\begin{equation}\label{Einsteinu_c}
\fom_{ab} +\eom^c\frac{\delta u_c}{\delta g^{ab}} =0.
\end{equation}
In some cases the Einstein-aether action is supplemented with a Lagrange multiplier term
enforcing the unit constraint,
\begin{equation}\label{Slambda}
S_\lambda = -\int d^4x\, \tilde\lambda\left(g_{ab} u^a u^b -1\right).
\end{equation}
This will contribute a $\tilde\lambda$-dependent term to the Einstein equation as well as 
the aether variations.  We will write such terms explicitly when they appear; the quantities $\eom_{ab}$, $\fom_{ab}$ and $\eom_a$ 
are always defined by \eqref{EabFab} and \eqref{Ea}, with
the action $S$ given by \eqref{eqn:action}.

Most of the theories considered in this paper involve scalar fields that determine
the background structures on which the theory is based.  We will often make use
of the fact that the scalar equations of motion are 
implied by the other field equations.  The proof
of this is straightforward: consider an action $S[g_{ab}, \chi, \Phi^i]$ that is a diffeomorphism
invariant functional of the metric, other tensor fields $\chi$, and scalar fields $\Phi^i$.
Under a diffeomorphism generated by $\xi^a$, the action varies by
\begin{equation}
\delta S = \int\left(\frac{\delta S}{\delta g^{ab}} \lie_\xi g^{ab} +\frac{\delta S}{\delta \chi}\lie_\xi \chi + \frac{\delta S}{\delta \Phi^i} \lie_\xi \Phi^i \right)
\end{equation}
This variation must vanish for all vectors $\xi^a$, and since
 the first two terms are zero when the metric and $\chi$ field equations hold, we find that
 \begin{equation}\label{dSdPhi}
 \frac{\delta S}{\delta \Phi^i} \nabla_a \Phi^i = 0.
 \end{equation}
 As long as the gradients $d\Phi^i$ are (non-vanishing and) linearly independent, 
 which can hold for up to four scalars, this implies that the scalar field 
 equations hold, $\delta S/ \delta \Phi^i=0$.
 If their equations of motion are automatic in this way, we can fix the scalars at the level 
 of the action without losing any dynamical information.  The gauge-fixed action
 is no longer  invariant under the full diffeomorphism group.
 In our application, where a vector field $u^a$ is constructed 
using scalars, the gradients of the scalars 
must be linearly independent 
in order for $u^a$ to be nonsingular and non-vanishing. Thus,
for the physically relevant configurations, those scalar field equations are automatic.\footnote{It can happen
that the gradients fail to be independent on a set of measure zero, e.g.\ on a codimension one surface, but 
with $u^a$ remaining well-defined in the limit as that surface is approached. In that case presumably continuity implies that the scalar field equations also hold directly on that surface, at least provided the fields are all nonsingular.}
A scalar Lagrange multiplier, on the other hand, need not have non-vanishing gradient, 
so its field equation (i.e.\ the corresponding constraint) should be imposed directly.

Finally, we recall that in any diffeomorphism invariant theory, some
of the field equations are constraints on  initial data, rather than
evolution equations.  For Einstein-aether theory, which contains a
dynamical, contravariant vector field $u^a$, the quantities
\begin{equation}\label{eqn:contraconstr}
\constr^{(t)}_b = \nabla_a t\left(2\eom\indices{^a_b} +u^a\eom_b\right).
\end{equation}
contain no more than first partial derivatives with respect
to $t$, for any choice of $t$ and the remaining three coordinates
\cite{Seifert2007a,Jacobson2011}.
When the field equations are satisfied, we have
$\eom\indices{^a_b}=-\tilde\lambda u^au_b$ and  $\eom_b=
2\tilde\lambda u_b$, where $\tilde\lambda$ terms arise from the
Lagrange multiplier term \eqref{Slambda}.  The $\tilde\lambda$ terms
thus cancel, so in Einstein-aether theory $\constr^{(t)}_b$ vanishes
on shell.
When $t$ is a time evolution coordinate, these constraints thus
restrict the allowed initial data.
For a covariant aether vector $u_a$, the expression for the constraint
has a different appearance,
\begin{equation}\label{eqn:covconstr}
\constr^{(t)}_b = \nabla_a t\left(2\fom\indices{^a_b}-\eom^a u_b\right),
\end{equation}
but is in fact the same as a consequence of \eqref{eqn:EFreln}.

More generally, in the various theories we consider here, although the
quantities \eqref{eqn:contraconstr} (or \eqref{eqn:covconstr}) will
have no higher than first $t$-derivatives (in appropriately adapted
gauges), some or all of them may not vanish when the field equations
hold, because the $u^a$ (or $u_a$) field equation per se is not
imposed. For this reason, we refer to them generally as ``constraint
quantities," rather than as ``constraints." This failure of constraint
equations to hold corresponds to the lack of diffeomorphism invariance
of the gauge-fixed action.  We shall analyze the form of the
constraint violation in each case as a means
of characterizing the extra freedom available in solutions to these theories.

\section{Twisting aether: threading theories} \label{sec:twisting}

A twisting aether flow does not determine a preferred foliation of spacetime
by spacelike hypersurfaces, but it does define a preferred  
threading of spacetime. 
In Einstein-aether theory, this threading is specified by an independently varied 
vector field modulo local scale, from which the aether 4-velocity 
$u^a$ \eqref{u(A)} 
is constructed with the use of the metric (alternatively, one can use a
Lagrange multiplier term to enforce the unit constraint on $u^a$).
In this section we consider three other ways of constructing $u^a$. 
In the first subsection, 
the threading is determined by three 
scalar fields which are Lagrangian (comoving) coordinates for the aether.
In the following two subsections, an additional scalar field $\psi$ is introduced into
both the line field and the Lagrangian coordinate constructions of $u^a$. 
The field $\psi$ is an independent ``clock" that marks time along the threads,
and is constrained to agree with proper time by a Lagrange multiplier term. 
These constructions are all very closely related to each other, but they 
yield theories that differ insofar as different integration constants are required
to determine a solution, corresponding to different initial value constraints that are violated.

\subsection{Fixed threading theory} \label{sec:bgthreads}
A threading can be specified as the curves along which three scalar fields 
$\vphi^I$, $I=1,2,3$ are all constant.  If the 
theory is to depend only on these curves as one dimensional sub-manifolds,  
and not on any parameterization, the action must be invariant under 
all smooth invertible field redefinitions of the scalars,
\beq\label{phidiff}
\vphi^I\mapsto\bar\vphi^I(\vphi^J).
\eeq
This can be achieved by restricting the action 
to depend on $\vphi^I$ only via the unit aether 4-velocity
\beq\label{u(At)}
u^a(\vphi^I,g_{ab}) =\Jt^a/|\Jt|,
\eeq
with $\Jt^a$ the metric-independent vector density 
\beq\label{Atilde}
\Jt^a = \et^{abcd}\vphi^1_{,b}\vphi^2_{,c}\vphi^3_{,d},
\eeq
where $\et^{abcd}$ is the alternating symbol, i.e.\ the Levi-Civita tensor density of weight 1.
The vector field defined in \eqref{u(At)} is invariant under the ``$\vphi$-diffeos" \eqref{phidiff}, 
since both the numerator and denominator are rescaled by the Jacobian determinant 
$\det\left({\partial\bar\vphi^I}/{\partial\vphi^J}\right)$. 
The corresponding action \eqref{eqn:action} is then a functional of the metric and 
the three scalar fields.

Note that the action 
is quadratic in second derivatives of the scalars $\vphi^I$.
This implies that the 
field equations will be fourth order in derivatives of the
scalars, and third order in derivatives of the metric (arising from the Christoffel
connection terms).  However, as explained in section \ref{sec:prelims}, we may treat the 
scalars as fixed, 
not varied in the action, without changing the 
dynamical content of the theory.  
Since the $\vphi^I$ define a threading, we
call this the \textit{fixed threading theory}.
In the co-moving gauge, where
$\vphi^I$ are equal to the spatial coordinates, $u^\alpha$ contains no derivatives, and 
the 
field equations arising from metric variations are of second order.

\subsubsection{Relation to zero temperature perfect fluid}

The dynamics of perfect fluids was formulated long ago in terms of 
three Euler potentials $\vphi^I$ \cite{Carter1972,Andersson2007}, a formulation that has recently 
been fruitfully exploited with the application of ideas from effective field theory (see e.g. \cite{Dubovsky2012}). 
In that setting, the vector density $\Jt^a$ represents the 
conserved entropy current and $|\Jt|$
is the entropy density in the fluid rest frame. The entropy current
is invariant under $\vphi$-diffeomorphisms with unit Jacobian determinant.
Unlike for our aethereal application, 
full $\vphi$-diffeo invariance is not imposed, because the entropy density 
is physically meaningful. The presence of 
 conserved particle number necessitates 
an additional scalar field 
with a shift symmetry in the action. Our 
``clock field" $\psi$ introduced below [see e.g.\ \eqref{eqn:threadparam}] is directly analogous to this, although
the corresponding chemical potential $u^a\nabla_a\psi$ 
is required by the unit norm constraint to be
everywhere equal to unity.

For fluids without conserved particle number,
the action 
at first order in derivatives is the integral of minus the 
energy density  expressed 
as a function $\rho(b)$ of the entropy density scalar
$b=|\Jt|/\sqrt{-g}$. 
The function $\rho(b)$
determines the equation of state of the fluid.
The aether fluid has the  
property that at first
derivative order its energy density is independent of the entropy density,
as required by the full $\vphi$-diffeo symmetry. This happens for
a thermal fluid only at zero temperature, hence the aether can be 
considered a \textit{zero temperature fluid}. 
The action for such a fluid is just proportional to the spacetime volume,  so
the stress-energy tensor at this derivative order is nothing but a 
(possibly vanishing) cosmological constant,
motivating the name ``vacuum fluid" for the aether.
The dynamics of the vacuum fluid is governed at lowest derivative order by 
the action \eqref{eqn:action} involving the ``strain" of the fluid.

\subsubsection{Comparison with Einstein-aether theory}

Under variations of the fields, 
in $u^a(A^m,g_{mn})$  \eqref{u(A)} and $u^a(\vphi^I, g_{mn})$ 
\eqref{u(At)}, 
the variation of $u^a$ has both a parallel and a
perpendicular part,  
\beq
\d u^a = \d u^a_\parallel + \d u^a_\perp.
\eeq
The metric variation generates only $\d u^a_\parallel$, while the
$A^m$ and $\vphi^I$ variations generate only $\d u^a_\perp$. 
The metric-induced variation in both cases 
is just what is needed to keep $u^a$ a unit vector:
$0=\d(g_{mn}u^m u^n) = (\d g_{mn})u^mu^n + 2 g_{ma} u^m \d u^a$ implies that 
$\d u^a_\parallel = -\tfrac12 u^a u^mu^n\d g_{mn}$, hence
\beq\label{deltauparallel}
\d u^a_\parallel = \frac12 u^a\, u_mu_n\d g^{mn}.
\eeq
The perpendicular part of the variation is given  in Einstein-aether theory by 
\beq\label{deltauperpEA}
\d u^a_\perp = (\d^a_m -u^au_m) \d A^m
\eeq
and in the fixed threading theory by 
\beq\label{deltauperp}
\d u^a_\perp = \frac1{2|\Jt|}(\d^a_m -u^au_m) \et^{mbcd}\e_{IJK}\vphi^I_{,b}\vphi^J_{,c}\,\d\vphi^K_{,d}.
\eeq
Thus the metric equation of motion \eqref{Einsteinu^c}
is the same
in terms of $g_{mn}$ and $u^m$ in the background threading theory
as it is in Einstein-aether theory,
\beq \label{eqn:EAeEE}
\eom_{ab}+\frac12u^c\eom_c u_a u_b = 0.
\eeq
The remaining equations of motion arise in both theories from 
the variation $\d u^a_\perp$, and here a discrepancy arises. 

In Einstein-aether theory the equation of motion arising from perpendicular aether variation
\eqref{deltauperpEA}  is
\beq\label{ueom}
\eom^\perp_m\equiv (\d^a_m -u^au_m)\eom_a=0,
\eeq
while in background threading theory it is 
\beq\label{phieom}
\eom_K\equiv \frac{\d S}{\d \vphi^K}=0.
\eeq
Diffeomorphism invariance of 
the action implies that the scalar equations 
\eqref{phieom} hold as a consequence of 
the Einstein equation [cf.\ discussion around Eq. \eqref{dSdPhi}],
so they add no new information.
On the other hand, the perpendicular aether equation \eqref{ueom} adds 
restrictions in Einstein-aether theory.

To discover the precise relation between the equations of motion
\eqref{ueom} and \eqref{phieom}, 
note that since 
$u^a(\vphi^I,g_{mn})$ is constructed covariantly,
diffeo variations of its arguments induce its diffeo variation
$\d u^a=\lie_\xi u^a$ as a vector field. 
In particular, the perpendicular component of $\lie_\xi u^a$ is equal to 
the variation induced via $\lie_\xi \vphi^K$, so the corresponding 
contributions to the variation of the action 
$S[g_{ab},u^a(\vphi^I,g_{mn})]$ are also equal,
\beq 
\int \eom^\perp_m \lie_\xi u^m =\int \eom_K \lie_\xi \vphi^K.
\eeq
Now using $\lie_\xi u^a = -\lie_u \xi^a$ and integrating by parts we obtain
\beq 
\int (\lie_u\eom^\perp_m - \eom_K \vphi^K_{,m})\xi^m=0
\eeq
for all vector fields $\xi^m$, which yields the identity
\beq\label{EuEvphi}
\lie_u\eom^\perp_m =\eom_K\, \vphi^K_{,m}.
\eeq
Thus \eqref{ueom} implies \eqref{phieom}
(provided again that the gradients $\vphi^K_{,m}$ are linearly independent), but 
\eqref{phieom} implies only that the Lie derivative of \eqref{ueom} holds.

We thus see that in the fixed threading theory the Einstein equation implies 
\beq\label{lieueomperp}
\lie_u\eom^\perp_m=0.
\eeq
Put differently, instead of the aether equation \eqref{ueom} one has
\begin{equation} \label{eqn:aethersource}
\eom^\perp_m= - {\tilde\mu}^\perp_m,
\end{equation}
where the ``source term" ${\tilde\mu}^\perp_m$ is a covector density 
that satisfies 
$u^m {\tilde\mu}^\perp_m =0$ and 
is conserved along the aether flow,
\beq\label{lieukappa}
\lie_u{\tilde\mu}^\perp_m = 0.
\eeq
A transparent way
to express the conservation law \eqref{lieukappa} is to 
use
adapted coordinates, $x^I = \vphi^I$, and to
choose $x^0=\tau$ with $u^a\t_{,a}=1$, 
so that the components of the aether 4-velocity 
are all constant, $u^\a = \d^\a_\t$.  Then 
the components of the Lie derivative are just the partial derivatives with respect to 
$\t$,  and  \eqref{lieukappa} takes the simple form
\beq \label{dkappa/dt}
\partial_\t  {\tilde\mu}^\perp_\a = 0.
\eeq
The three components 
${\tilde\mu}^\perp_I$ are then just constants of integration on each thread,
while ${\tilde\mu}^\perp_\t$ vanishes identically.
The freedom to choose these integration constants different from 
zero is what distinguishes the fixed threading theory from Einstein-aether theory.
For lack of a better name, we shall call ${\tilde\mu}^\perp_m$ the \textit{vector source density}
(VSD).

The identity \eqref{EuEvphi} shows that the $\vphi^K$ equation of motion is
``weaker" than the perpendicular $u^m$ equation of motion, but this discrepancy 
remains a bit mysterious, since it would seem that variations of $\vphi^K$ produce
all possible perpendicular variations of $u^m$. Of course the difference must 
arise because $\vphi^K$ occurs in the action with an extra derivative, but why 
exactly is that important? The answer lies in the boundary conditions. When we 
drop boundary terms we are holding $\vphi^K$  fixed at the boundaries, in particular
the initial and final boundary. This entails an integral constraint 
on the $u^m$ variations (the endpoints of each thread are fixed),  
which translates into the fact that the
$\vphi^K$ variations imply only the time derivative of the $u^m$ equation of motion. 

A simple example serves to illustrate this point. Suppose we have a mechanical 
system in one dimension with Lagrangian $L(x,\dot{x})$, 
and we make the replacement $x = \dot{y}$, and treat $y(t)$ as the basic 
dynamical variable.
Then the action variation is
$\d S = \int (\d S/\d x) \d \dot{y} = -\int (d/dt)(\d S/\d x) \d y$, so the $y$ equation 
of motion is the time derivative of the $x$ equation of motion. 
It is weaker than the $x$ equation of motion because not all $x$ variations are
included in the $y$-version of Hamilton's principle. 
Since the initial and final values $y_{1,2}$ are fixed in the $y$-variations, there is an
implicit constraint on the integral of $x$ due to the fact that 
$\int x \, dt = \int\dot{y}\, dt = y_2-y_1$.  We could include this constraint directly in
the $x$-version of Hamilton's principle with the addition of a Lagrange multiplier 
term $\l(\int x\,dt-\Delta y)$. The result would be the equation of motion
$\d S/\d x=\l$, where the Lagrange multiplier $\l$ is an undetermined constant
corresponding to a constant external force.  In the $y$ equation, $\lambda$ 
corresponds to the extra integration constant needed to specify a solution.

How does a nonzero  
VSD for the aether field equation
change the aether theory?
In another paper, we find that it does not alter the Newtonian limit or static, spherical
stars (assuming no radial aether component) \cite{Jacobson2015},
and by symmetry homogeneous, isotropic cosmology is unaltered.  
However, it acts as an external force for wave 
modes, shifting the equilibrium amplitude away from zero. 
In the next subsection we show that, more generally,
the source density integration constants 
characterize a violation of the initial value constraints of 
Einstein-aether theory.
In the following subsection we show that 
magnitude of the source density is diluted 
as the aether expands with the universe, which
suppresses its observable consequences.

\subsubsection{Initial value constraint violation}  \label{sec:ivcthreading}

The VSD $\tilde\mu_m^\perp$  
in (\ref{eqn:aethersource}) suggests that the 
fixed threading theory requires more initial data than Einstein aether theory, since  
$\tilde\mu^\perp_m$ is a freely specifiable initial source for the aether equation.  
This new freedom can be characterized in terms of violated Einstein-aether initial value 
constraints.  

For an arbitrary fourth coordinate $x^0$, the constraint quantities
 \eqref{eqn:contraconstr} take the form 
\begin{equation}\label{eqn:constraint}
\constr_\a^{(0)} = 2\eom\indices{^0_\a}+u^0\eom_\a,
\end{equation}
When the metric field equation \eqref{eqn:EAeEE} holds, 
\eqref{eqn:aethersource} implies that these quantities
are nonvanishing and instead satisfy
\begin{equation}\label{random}
\constr^{(0)}_\a =  -u^0  {\tilde\mu}^\perp_\a.
\end{equation}
The 
$u^a$ component constraint $u^\a\tilde\constr_\a^{(0)}=0$ holds, 
since $u^\a{\tilde\mu}^\perp_\a=0$, but the 
three ``perpendicular constraints" are violated.

In adapted coordinates, $x^I = \vphi^I$,
the constraint violation is preserved in time. 
This is easiest to see with the choice $x^0=\t$, with
$\tau$ the proper time along the threads.
Then we have $u^0=1$, and 
\eqref{dkappa/dt} shows that the components of the 
constraint $\tilde\constr^{\scriptscriptstyle(\t)}_\a$ are preserved in $\t$.
In fact the same result holds for any choice of the fourth coordinate
$x^0$: the condition $u^\a{\tilde\mu}^\perp_\a=0$ implies that under a change
from $\t$ to $x^0$, the components of the covector density ${\tilde\mu}^\perp_\a$
change only by the Jacobian factor $\partial\t/\partial x^0$, while $u^0 = (\partial x^0/\partial \t)u^\t$. Therefore
the components of the $x^0$-constraint \eqref{random} in $(x^0,x^I)$ coordinates are the
same as those of the $\t$-constraint in $(\t,x^I)$ coordinates.  Since 
$u^0\partial_0 = \partial_\t$, the previous result implies that also\footnote{A 
covariant version of this argument uses  \eqref{lieukappa} and 
the identity $\lie_{u/(u\cdot dx^0)}[(u\cdot dx^0){\tilde\mu}^\perp_m] = \lie_{u}{\tilde\mu}^\perp_m$,
which holds in view of the unit density weight and $u^m\tilde\mu^\perp_m=0$.}  
\begin{equation}\label{eqn:constrpreserve}
\partial_0\constr_\a^{(0)}= 0.
\end{equation}
This equation shows that the new freedom takes the form of an infinite 
collection of conserved quantities. 
The constraint violation may be freely specified 
at an initial time, but remains constant at all subsequent times.

The vanishing of the constraint quantities in Einstein-aether theory is a consequence
of full spacetime diffeomorphism symmetry. The fixed threading theory respects only the
thread preserving diffeomorphisms, which in adapted coordinates take the form
$t\mapsto \bar{t}(t,x^I)$,  $x^I\mapsto \bar{x}^I(x^J)$. Intuitively, 
since we cannot perform arbitrary gauge transformations of the spatial coordinates
as we evolve in time, 
there should be no constraints associated with those diffeomorphisms imposed
on the dynamics.
This is why we find that $\constr^{(0)}_I$,  the spatial constraint quantities for 
evolution along the threads, are non-vanishing.

By contrast, for evolution with respect to a parameter
that is constant on the threads, say $x^3$, 
all constraint quantities vanish, since $u^a\nabla_a x^3=0$, so the number of initial value constraints
remains equal to four. This might be expected
since as we evolve in $x^3$, we can perform both time and spatial diffeomorphisms.
(That these are required to preserve the fibers evidently does not cause the constraints to be lost.)
This gauge symmetry means the dynamics cannot be fully deterministic, so that some
field equations must be constraints.

\subsubsection{Cosmological evolution of source density}
In homogeneous isotropic symmetry, the VSD necessarily 
vanishes, and the background threading theory is
identical to Einstein-aether theory.  It is natural to imagine
some kind of fluctuations around the symmetric configuration however. Since 
the VSD arises as integration constants, its power spectrum cannot be derived 
from the properties of quantum vacuum fluctuations. At this 
point we have identified no principle to select a primordial spectrum of VSD.
What we can say however is that the amplitude will decrease as
the universe expands. 

To characterize the amplitude of the VSD we use the scalar quantity 
\beq\label{kappa}
\kappa \equiv [g^{ab}{\tilde\mu}^\perp_a {\tilde\mu}^\perp_b/(-g)]^{1/2}.
\eeq
An approximate redshift law for $\k$ can be easily obtained by using
for $g_{ab}$ the homogeneous isotropic metric $ds^2 = dt^2 - a(t)^2 dx^i dx^i$, 
and neglecting the anisotropic corrections to the conservation law 
\eqref{lieukappa}. Then the coordinates
$(x^i,t)$ are adapted to $u^a$, and the conservation law takes the form
$\partial_t {\tilde\mu}^\perp_i=0$, so \eqref{kappa} yields $\k\propto a^{-4}$. The physical
effects of the VSD therefore decrease like those of radiation as the universe expands.

\subsection{Fixed clock theory: a ponderable aether}\label{sec:clockthreads}

In section \ref{sec:bgthreads}, we introduced three scalar fields that defined a threading.  In the co-moving gauge, these scalars have the effect of breaking spatial diffeomorphism symmetry
when fixed at the level of the action.   
In this section we consider a different theory, in which temporal rather than spatial 
diffeomorphism symmetry is broken.
This involves introducing a 
``clock" field $\psi$ that defines a preferred notion of time along 
the aether flow.

Since the clock field $\psi$ is a scalar, we may again fix $\psi$ to a background value 
at the level of the action.  In analogy to the fixed threading we expect this fixed clock to 
lead to a violation of an initial value constraint and therefore to produce, 
in effect, an additional
degree of freedom.  The constraint violation in this case is quite analogous to
the ``dark matter as an integration constant" \cite{Mukohyama2009}
in projectable \Horava{} gravity.
The latter is due to the absence of the local Hamiltonian constraint 
in that theory. That constraint normally arises from the variation of the lapse function
$N=(g^{tt})^{-1/2}$, but in projectable  \Horava{} gravity $N=N(t)$ depends only on $t$.
It is therefore not varied independently at each point on a constant $t$ surface, so the 
associated local constraint is not imposed.  
The covariant construction of an aether with a fixed clock given here
yields a similar effect. Unlike in the projectable \Horava{} case, however,
the ``dark matter mass current" is conserved in the fixed clock aether theory.

We start
as in 
Einstein-aether theory with a dynamical vector field $A^a$ 
but, rather than defining the aether 4-velocity dividing by $|A|$,
we 
define it by
\beq\label{eqn:pondu}
u^a(A^m,\psi)=\frac{{A}^a}{{A}^m\psi_{,m}},
\eeq
where $\psi$ is the clock field. 
By construction we have $u^a \psi_{,a}=1$, so $\psi$ is a parameter 
on the aether flow compatible with $u^a$. 
Note that (\ref{eqn:pondu}) is unchanged under a thread-dependent shift
$\psi \mapsto \psi + \upsilon$, with $\upsilon$ constant along each thread, 
$A^a\nabla_a\upsilon = 0$ (note this symmetry was called a ``chemical shift'' 
in the works on effective field theory for fluids \cite{Dubovsky2012}). 
The requirement of this symmetry precludes standard
kinetic or potential terms for $\psi$ and, since $u^a\psi_{,a}=1$,
a term like $(u^a\psi_{,a})^2$ only adds a constant to the action.

Unlike its Einstein-aether cousin 
$u^a(A^m,g_{mn})$  \eqref{u(A)},
$u^a(A^m,\psi)$ is not a unit vector by construction, 
so we impose the unit constraint by adding a Lagrange
multiplier term \eqref{Slambda} to the action,
enforcing the relation
\beq\label{A.dpsi}
(A^m\psi_{,m})^2 =  g_{mn}A^m A^n.
\eeq
It seems at first that this
could be satisfied either by solving a first order ODE for $\psi$
on each thread, or by restricting $g_{mn}$ (the condition is independent of the scale of 
$A^m$ so it can not be satisfied by restricting that scale). However, 
solving (\ref{A.dpsi}) for $\psi$ by integrating along each thread
would be inconsistent with fixing $\psi$ at both endpoints in Hamilton's principle 
unless further constraints on variations of $A^a$ and $g_{ab}$ are imposed.
Instead, it is simplest to view the unit constraint as fixing a component of the metric
 in terms of $A^m$ and $\psi_{,m}$.  
Since $\psi$ is a scalar field, its equation of motion is satisfied by virtue of the other 
equations of motion
(provided $\psi_{,m}\ne0$).
It can therefore be considered fixed.
We call this the \textit{fixed clock theory} since, 
when the unit constraint is satisfied, 
$\psi$ marks proper time on each thread. 

The variation of \eqref{eqn:pondu} is given by 
\beq\label{deltauFP}
\d u^a = -u^a u^m \d\psi_{,m}+ \frac{1}{A\cdot d\psi}(\d^a_m - u^a \psi_{,m})\d A^m.
\eeq
The $\psi $ equation of motion thus takes the form of a current conservation law,
\beq\label{mueqn}
(\tilde\mu u^m)_{,m}=0,\qquad \tilde\mu \equiv \mu \sqrt{-g}\equiv u^a\left(2\tilde\lambda u_a-\eom_a \right),
\eeq
The $A^a$ equation of motion is
\beq 
\frac{1}{A\cdot d\psi}(\d^a_m - u^a \psi_{,m})(\eom_a-2\tilde\lambda u_a)=0,
\eeq
which is equivalent to 
\beq\label{ueqnpsi}
\eom_a-2\tilde\lambda u_a =-\tilde\mu \psi_{,a}.
\eeq
In this theory, $u^a$ has no metric dependence, so the Einstein equation is
\beq\label{geqnpsi}
\eom_{ab} +\tilde\lambda u_a u_b = 0,
\eeq
which in light of \eqref{mueqn} becomes
\beq
\eom_{ab}+\frac12 u^c\eom_c u_a u_b +\frac12\tilde\mu u_a u_b = 0.
\eeq
The aether stress tensor 
thus picks up the extra dust-like contribution, $\mu u_a u_b$,
which is not present in Einstein-aether theory.  
Together with the conservation equation \eqref{mueqn}
this suggests the interpretation of 
$\tilde\mu$ as the internal energy density of the aether, 
and motivates the descriptive term \textit{ponderable aether}.
Notice that even in the absence of the aether terms in the 
action \eqref{eqn:action}, the Lagrange multiplier term \eqref{Slambda} alone
suffices to introduce the aethereal dust stress tensor $\mu u_a u_b$.

The sign of $\mu$ is not fixed, so the aethereal dust
can contribute negative energy density to the source in the 
metric field equation.
Normally the presence of negative energy could produce an instability, since the system 
could evolve to a highly excited state of compensating
positive and negative energy components. However,
as a consequence of the conservation equation (\ref{mueqn}), 
the spatial integral of $\mu$ is conserved along any bundle of aether 
worldlines, precluding instabilities of that sort. 

The initial value formulation of the fixed clock theory 
differs from that of Einstein-aether theory
by a violated constraint equation.
As explained previously for the fixed threading theory, 
the quantity $\constr_\a^{(t)}$ defined in \eqref{eqn:contraconstr}
has only first $t$-derivatives  
of the metric and the aether 4-velocity. 
For the background clock theory it has second derivatives, since 
the aether 4-velocity \eqref{eqn:pondu} involves the derivative of 
$\psi$. In the adapted coordinate ``clock gauge" $x^0=\psi$, however, $u^\a$ is
algebraic,
\beq
u^\a(A^\mu,\psi)=A^\a/A^0,
\eeq
so $\tilde\constr_\a^{(t)}$ has only first $t$-derivatives for any choice of $t$.
When the $A^a$ and metric equations \eqref{ueqnpsi} and \eqref{geqnpsi}
hold, we have
\beq\label{Ctpsi}
\constr_b^{(t)}= -(u^a t_{,a})\tilde\mu \psi_{,b}.
\eeq
The right hand side vanishes when contracted with any vector tangent to the
constant $\psi$ surface, so the presence of nonzero $\tilde\mu$
in the aether equation (\ref{ueqnpsi}) leads to a single constraint violation.
The additional freedom in the theory is parameterized by $\tilde\mu$. 
If we choose $t=\psi$ as the evolution parameter 
and use the clock gauge, \eqref{Ctpsi} takes the form 
\begin{equation}\label{eqn:FPconstraint}
\constr_\a^{(\psi)} =-\tilde\mu \d^0_\a.
\end{equation}
If we further choose spatially adapted coordinates, so that 
$u^\a=\delta^\alpha_0$, the right hand side of equation (\ref{eqn:FPconstraint}) is constant in 
$x^0$ as a consequence of \eqref{mueqn}.  As in the fixed threading theory, 
the new degrees of freedom are, in this sense, ``totally integrable."

We have found that for evolution with respect to any coordinate 
$t$ such that $u\cdot dt\neq0$, the $\psi$-component
of the constraint quantities does not vanish.  This is because the fixed clock
breaks time diffeomorphism symmetry.  The clock shift symmetry remains,
but it allows for only a single, time independent shift, so 
the $\psi$ surfaces cannot be deformed as we evolve along the threads.
The components of the constraints in the directions tangent to the 
$\psi$ surfaces are preserved, since the threads
are determined by a dynamical vector field, rather than by a background
structure.

The contribution $\mu u_a u_b$ has the form of a pressureless 
fluid source in the Einstein equation, but its divergence is 
not zero when the aether is not geodesic. 
In homogeneous isotropic cosmology, however, it does behave exactly
as pressureless dust, with $\mu \propto a^{-3}$.  
During an inflationary period $\mu$ would be exponentially
suppressed, so in the standard inflationary cosmological model it would 
presumably be too small today to have any observable effect. 
If there were some way to transcend the classical conservation law for 
$\mu$ and generate a nonzero  value around the time of matter radiation
equality, it could play the role of the homogeneous dark matter in a 
$\Lambda$CDM model.\footnote{If it were generated earlier, e.g.\ at reheating,
it would quickly dominate unless fine-tuned to an extremely 
small value relative to the radiation energy density.  Mechanisms for 
generating dark matter after inflation have been proposed for the related 
projectable \Horava gravity \cite{Mukohyama2009} and mimetic dark matter 
\cite{Chamseddine2013, Mirzagholi2014} theories.}
This leads to the question of 
how it would behave as structure forms. Its nongeodesic character 
suggests that it would not form structure in the manner of geodesic 
dark matter. In another paper \cite{Jacobson2015}
we have examined the growth of linearized perturbations, 
and found that if $\mu$ were to comprise the homogeneous dark matter
density at early times it would lead to an unacceptably high
growth rate on super-horizon scales and no growth on sub-horizon scales.

\subsection{Fixed aether theory}\label{sec:bgaether}

In the previous two sections, we considered a theory with broken spatial 
diffeomorphisms, the fixed threading, and one with broken temporal diffeomorphisms,
the fixed clock.  In this section we  combine these features
and consider an aether theory with broken
spatial and temporal diffeomorphisms. 

We now define the aether 4-velocity by 
\beq\label{eqn:threadparam}
u^a(\vphi^I,\psi)=\frac{\Jt^a}{\Jt^m\psi_{,m}},
\eeq
where $\Jt^a$ is the vector density constructed from $\vphi^I$ defined in equation (\ref{Atilde}),
and $\psi$ is a scalar clock field. 
Like the aether 4-velocity of the background threading theory \eqref{u(At)}, 
$u^a(\vphi^I,\psi)$ is unchanged under all $\vphi$-diffeos \eqref{phidiff} and,
as in the fixed clock theory, it has 
the clock shift symmetry,
 $\psi\mapsto\psi+\upsilon(\vphi^I)$ (again, this corresponds to the ``chemical shift''
 in the effective theory of fluid dynamics \cite{Dubovsky2012}).   
 Also, as in the latter theory, 
it is not normalized
by construction,
so the Lagrange multiplier term \eqref{Slambda} is 
again used to impose the unit constraint.

Since $u^a(\vphi^I,\psi)$ is constructed entirely from four scalar fields,
its variations arise solely from variations of the four scalars. As explained
above, provided the
four gradients $\vphi^I_{,a}$, and $\psi_{,a}$ are linearly independent, which is required
for them to define a threading with parameter $\psi$, the equations of motion
for the four scalars will follow from the Einstein equation. 
They can therefore be held fixed in the action.
This \emph{fixed aether theory} is thus equivalent to Einstein-aether theory with an
aether vector $u^a$ that is not varied in the action; 
that is, the aether field equation
$\eom_a$ of Einstein-aether theory is not imposed. 

When the $g_{ab}$ and $\tilde\lambda$ equations of motion hold, however, 
the (vanishing) variation of the action $S[g_{ab},u^a,\tilde\l]$ with respect to a  
diffeomorphism generated by $\xi^a$ is given by 
$\int (\eom_a-2\tilde\lambda u_a) \lie_\xi u^a = \int \xi^a \lie_u( \eom_a-2\tilde\lambda u_a)$.
Since this vanishes for all $\xi^a$ we infer that 
\beq
\lie_u ( \eom_a-2\tilde\lambda u_a)=0.
\eeq
This is similar to \eqref{lieueomperp} in the fixed threading theory, 
but includes the component of 
$\eom_a$ along $u_a$. Thus although
the aether field equation is not imposed, it holds with the addition of an undetermined,
``constant" source term,
\begin{equation}\label{eqn:ponderable}
\eom_a -2\tilde\lambda u_a=-\tilde\mu_a,\qquad \lie_u\tilde\mu_a = 0.
\end{equation}
{Equation (\ref{eqn:ponderable}) implies} 
\beq\label{lambda}
\tilde\lambda=\frac12\left(\tilde\mu+ u^a \eom_a\right),
\eeq
with
\beq
\tilde\mu\equiv\mu \sqrt{-g}\equiv u^a \tilde\mu_a.
\eeq
Therefore, as in the fixed clock theory, 
the aether stress tensor 
contribution $(\tilde \lambda/\sqrt{-g}) u_a u_b$ 
picks up the extra term $\mu u_a u_b$
not present in Einstein-aether 
theory. 
Also, \eqref{eqn:ponderable} and $\lie_u u^a=0$
imply  $\lie_u \tilde\mu=(\tilde\mu u^a)_{,a}=0$,
so that
$\mu$ acts like a ``dark matter'' source of gravity
that can be interpreted as the internal energy density 
of a ponderable aether. 

When we work in co-moving, clock gauge ($x^I=\vphi^I, x^0 = \psi$), 
the diffeomorphism symmetry is broken down to time independent transformations
$x^I\rightarrow f^I(x^J)$ and $x^0\rightarrow x^0 + f(x^J)$, so
we should expect all four constraints to be 
violated.
When the metric and $\tilde\lambda$ equations are satisfied, 
the $x^0$-constraint quantity \eqref{eqn:contraconstr} 
for the fixed aether theory in these coordinates takes the form
\begin{equation}\label{eqn:FEconstraint}
\constr_\a^{(0)} = 2\eom\indices{^0_\a}+u^0\eom_\a =-\tilde\mu_\a.
\end{equation}
This indeed confirms that all four initial value constraints 
of Einstein-aether theory are  violated.  
This is as expected, since for evolution with respect to any parameter that
advances along the threads,
there remains no diffeomorphism freedom that would make the dynamics
underdetermined.  For evolution with respect to a parameter $s$ that
is constant along 
the threads, we again find that 
all constraints vanish. In addition to the $s$-dependent thread preserving diffeomorphisms, 
the clock field's shift symmetry allows for 
$s$-dependent changes in $\psi$.
Thus, we find four additional 
initial value freedoms per spatial point, 
which again by \eqref{eqn:ponderable} are ``totally integrable."

\section{Foliation theories} \label{sec:foliation}
We now turn to theories 
involving a foliation of spacetime by
spacelike hypersurfaces.  These theories are distinct from  threading
theories because the aether vector, constructed as the unit normal to the foliation,
is necessarily twist-free.  The 
simplest
foliation type theory is  
khronometric theory, 
the low energy limit of nonprojectable \Horava gravity.  After reviewing its construction in section
\ref{sec:nonprojectable}, we proceed in section \ref{sec:clockfoliation}
to add a fixed clock to the theory as was done for 
the threading theories in section \ref{sec:clockthreads}.  
As in that case, the resulting theory exhibits 
constraint violation, and similarly contains
a ``dark matter'' component in the Einstein equation.  Finally, in section
\ref{sec:projectable} we consider
the case where the foliation and the clock field  
coincide.  This 
results in the projectable version of \Horava gravity, and we discuss
the  
relation between the constraint violation and
the ``dark matter as an integration constant'' 
of that theory \cite{Mukohyama2009}.

\subsection{Khronometric theory} \label{sec:nonprojectable}

A twist-free aether can be described by a scalar field $T$, 
dubbed the ``khronon," whose level sets define the hypersurfaces 
orthogonal to the aether 4-velocity \cite{Blas2009, Blas2010, Jacobson2010a}. 
In order that the theory 
depend only on the foliation by hypersurfaces, and not the values of $T$, 
the the action must be invariant under monotonic reparametrizations 
\beq \label{T-reparam}
T\rightarrow \bar{T}(T).
\eeq
The gradient of $T$ transforms as $\bar{T}_{,a} = (d\bar{T}/dT)T_{,a}$,
so the numerator and denominator of the aether 4-velocity \eqref{u(T)}
both acquire a factor $d\bar{T}/dT$, and these factors cancel.
Therefore the action $S[g_{ab}, u^a(T,g_{mn})]$  \eqref{eqn:action}
is invariant under $T$ reparametrizations.

Just as in the threading theory, this action is quadratic in \textit{second derivatives} of $T$,
so when $T$ is varied it yields equations of motion that are fourth order in 
derivatives of $T$ and third order in derivatives of the metric.
Again, as explained in section \ref{sec:prelims}, we may fix $T$ at the level
of the action without changing the dynamics.  In  
the adapted
gauge where 
$T$ is identified with one of the spacetime coordinates, 
we have from \eqref{u(T)}%
\beq
u^\a(T,g_{ab}) = g^{\a T}/\sqrt{g^{TT}}.
\eeq
Since the aether 4-velocity is an algebraic function of the metric components in this gauge,
and the action \eqref{eqn:action}
produces
a second order field equation for the metric
(terms with more than two \textit{spatial} derivatives occur in the full
\Horava{}-Lifshitz theory).
In this 
formulation, which is equivalent to 
\Horava's original one \cite{Horava2009}, 
the action is invariant only under $T$-foliation preserving diffeomorphisms,
together with $T$-reparametrizations \eqref{T-reparam}.

We can now examine the constraints for this theory.  The metric dependence 
of $u_c$ induces a variation $\delta u_c = -\frac12 u_c u_a u_b \delta g^{ab}$, so the
metric field equation \eqref{Einsteinu_c} reads
\begin{equation}\label{eqn:foliationEE}
\fom_{ab}-\frac12\eom^cu_c u_a u_b = 0.
\end{equation}
When this equation is satisfied, the constraint \eqref{eqn:covconstr} is equal to
\begin{equation}
\constr^{(t)}_b = -(\nabla_a t )\eom^a_\perp u_b.
\end{equation}
If we choose $t=T$ then, since 
$dT\propto u$, 
the right hand side vanishes, so
all the constraints $\constr^{(T)}_b$ vanish in 
the adapted gauge.

This is complementary to the situation described in 
section \ref{sec:ivcthreading} for the threading theory.  There we found that 
for evolution with respect to a parameter constant on 
the threads, the constraint quantities vanish.  He we find
a similar
result: for evolution 
with respect to a parameter $s$  that is constant on
the foliation, the constraint quantities vanish. 
We can still make 
$s$-dependent
time reparameterizations and spatial diffeomorphisms under 
this evolution, so we 
expect constraints associated with these gauge
transformations.  If instead we were to consider evolution 
with respect to a different parameter $s'$ that is \emph{not} constant on the foliation, 
we would find the $T$-component of the constraint violated.  This is
because the foliation cannot be deformed in an $s'$-dependent fashion, 
so the theory loses that gauge symmetry and the associated constraint.

\subsection{Fixed clock foliation theory}\label{sec:clockfoliation}

We can add a fixed clock to the foliation theory by following the 
method introduced above for the twisting aether.
We introduce the 
clock field $\psi$ and define the 
aether 4-velocity covector as  
\beq\label{uTpsi}
u_a(T,g_{mn},\psi)=\frac{T_{,a}}{g^{mn}T_{,m}\psi_{,n}},
\eeq
which is constrained by a Lagrange multiplier term to have unit norm.
The unit constraint requires that the lapse $N=(g^{ab}T_{,a}T_{,b})^{-1/2}$
be equal to $(g^{ab}T_{,a}\psi_{,b})^{-1}$, which freezes one metric degree of freedom
(in the adapted gauge, it fixes $(g^{T\psi})^2=g^{TT}$).
The field equations for both of the scalars $T$ and $\psi$ again follow from the 
Einstein equation, provided $T_{,a}$ and $\psi_{,a}$ are linearly independent.
Although the gradients $dT$ and $d\psi$ are both timelike, they will generically 
be independent whenever the aether is accelerated, since 
\begin{equation}
\lie_u (u_{[a}\psi_{,b]}) = a_{[a}\psi_{,b]}.
\end{equation}
Thus, the gauge in which both $T$ and $\psi$ are set equal 
to coordinates will generically be nonsingular.

The $\psi$ field equation gives a  conservation law, corresponding to 
its shift symmetry $\psi\rightarrow \psi +\text{const.}$,
\beq\label{eqn:muconserve}
[(\eom^a-2\tilde\lambda u^a) u_au^m]_{,m}=0.
\eeq
Defining, as usual, the scalar density $\tilde\mu=(2\tilde\lambda u^a-\eom^a) u_a$, 
we now have
\beq\label{lambdaPF}
\tilde\lambda=\frac12 \left( \tilde\mu+u_a \eom^a\right),
\eeq
as in the fixed aether 
case \eqref{lambda}.
The metric equation of motion receives a contribution from the 
metric variation in $u_a$, 
namely 
$\delta u_c = -u_c u_{(a}\psi_{,b)}\delta g^{ab}$.
The metric field equation is then
\begin{equation}
\fom_{ab} -\tilde\lambda u_a u_b-(\eom^c-2\tilde\lambda u^c)u_c u_{(a}\psi_{,b)} = 0.
\end{equation}
Rearranging to compare with (\ref{eqn:foliationEE}), this becomes
\begin{equation} \label{eqn:clockfoliationEE}
\fom_{ab}-\frac12\eom^cu_c u_a u_b +\frac12 \tilde\mu\left(u_a u_b +  2u_{(a}\psi_{,b)}^\perp\right)=0,
\end{equation}
where $\psi_{,b}^\perp=\psi_{,b}-u_b$ is the projection of $\psi_{,b}$ 
perpendicular to $u_b$.
The $\tilde\mu$ term gives the difference between this theory
 and the khronometric theory, and it takes the form of
non-geodesic ``dark matter" with 
momentum density.

Unlike the previous  
fixed clock theories,
here the 
clock field itself has an effect on the dynamics
via the perpendicular component of
$\psi_{,b}$ in the stress tensor \eqref{eqn:clockfoliationEE}. 
Since $u^m\psi_{,m}=1$,
$\psi$ is determined on each thread by its value at one point. Hence
the value of $\psi$ on one spacelike hypersurface must be 
chosen as initial data in order to integrate the equations of motion.

We now examine the constraints. Using $T$ as the time coordinate,  
enforcing the Einstein equation \eqref{eqn:clockfoliationEE},  
and using the fact that $u\propto dT$, we find
for the constraint quantities
\begin{equation}
\constr^{(T)}_b = -N^{-1} \tilde\mu \psi_{,b}.
\end{equation}
Thus the $T$-surface constraint is violated 
in the $\psi$-component, for essentially the same reasons given 
in section \ref{sec:clockthreads} for the fixed clock theory.
The conservation law \eqref{eqn:muconserve} 
implies $\lie_u\tilde\mu=0$, and we have $\lie_ud\psi=0$, so 
the constraint violation satisfies
\begin{equation}\label{eqn:lieNCbT}
\lie_u\left(N \constr_b^{(T)}\right) = 0.
\end{equation} 
This conservation law is more complicated 
 than the analogous one for the threading theory 
\eqref{eqn:constrpreserve}.  In the adapted gauge where $T$ and $\psi$ are coordinates
(and in which the field equations are second order in derivatives), the vector $u^a$ cannot
be chosen to be proportional to $\partial_T$, since $u\cdot d\psi = 1$.  Instead, we can
choose a gauge where $u = N^{-1} \partial_T +\partial_\psi$, in which case 
\eqref{eqn:lieNCbT} becomes
\beq
\partial_T \constr_\beta +  \partial_\psi (N\constr_\beta)  = 0.
\eeq
Hence, we see that the constraint violation evolves according to a first order 
differential equation that also involves $\psi$-derivatives of the metric component $N$.

\subsection{Projectable \Horava gravity}\label{sec:projectable}

Our final example of a foliation theory is projectable \Horava gravity, 
which can be obtained from khronometric gravity 
(section \ref{sec:nonprojectable}) by imposing
in the action the
restriction that the lapse
function be constant on each foliation surface, $N = N(T)$. 
The aether then takes the form $u_a=N(T)T_{,a}$.
An aether of this form satisfying the unit constraint is 
geodesic ($u^a\nabla_a u_b = u^a\nabla_b u_a = \half\nabla_b(u^a u_a)=0$), 
so the acceleration term $c_a a^2$ in the action (\ref{eqn:action}) becomes
superfluous.

The projectability restriction can be implemented by adding to the Lagrangian density
a Lagrange multiplier term $\tilde\lambda^{ab} N_{,a}T_{,b}$. 
Here $\tilde\lambda^{ab}$ is an antisymmetric tensor density, and 
$N(g,T)=(g^{mn}T_{,m}T_{,n})^{-1/2}$ is the lapse function. 
In the adapted gauge $x^0=T $, the new constraint term becomes 
$\tilde\lambda^{i0} N_{,i}$, with $N=(g^{00})^{-1/2}$. 
The Hamiltonian constraint, which arises from variations of the lapse,
now contains an additional piece $-\partial_i \tilde\lambda^{i0}$ from the 
Lagrange multiplier term.  This term represents a local violation of the 
Hamiltonian constraint.  Since the violation is a spatial divergence,
the integral of the Hamiltonian constraint will still be imposed
on a compact space without boundary.  This 
integrated constraint is the generator of global time reparameterizations,
and follows from the global variations of the lapse function $N(T)$ 
in the original $3+1$ formulation of projectable \Horava gravity.
With asymptotically flat boundary conditions, the integrated Hamiltonian 
constraint will not vanish, but equal the flux of $\tilde\lambda^{i0}$ through
the sphere at spatial infinity.  This is consistent with the fact that the metric 
is fixed at infinity, so, in particular, the variational principle does not include
global variations of the lapse.  

An alternative version of the projectable theory lacks the global
time reparameterization gauge symmetry, and correspondingly the
global Hamiltonian constraint is not imposed \cite{Horava2010,Xu2010,Anderson2012}.  
This version arises from a fixed clock foliation theory (section \ref{sec:clockfoliation})
when the clock is required in the action 
to be constant on the foliation slices.  To implement this we may add to the Lagrangian
a Lagrange multiplier term $\tilde\lambda^{ab}\psi_{,a}T_{,b}$, with 
$\tilde\lambda^{ab}$ an antisymmetric tensor density.  The corresponding 
constraint implies that $d\psi\propto dT$, so that $u^a$ \eqref{uTpsi} takes the form 
$u_a=\psi_{,a}/|d\psi|^2$.   
Having imposed this form for $u_a$, the $\tilde\lambda^{ab}$ constraint
can be omitted. At this point 
$T$ has disappeared from the action, 
its role being 
taken over by $\psi$.\footnote{We could have kept the role of $T$ explicit, 
using $d\psi\propto dT$ to express \eqref{uTpsi} as $u_a = T_{,a}/(\psi'|dT|^2)$. 
The unit constraint  then implies $\psi'|dT|=1$, so that we have $u_a=\psi'(T)T_{,a}$. The 
lapse function is thus given by $N=N(T)=\psi'(T)$.}
The unit constraint then implies that $|d\psi|=1$, so that 
$u_a=\psi_{,a}$,
and the lapse in the adapted gauge $x^0=\psi $
is fixed equal to 1.  There is no global lapse variation,
and no global Hamiltonian constraint.
This covariant formulation of projectable \Horava gravity 
has been described before in \cite{Blas2009, Jacobson2014}.\footnote{For some 
closely related theories, see \cite{Lim2010, Chamseddine2013, Haghani2014a,
Capela2014, Mirzagholi2014}.}

The Einstein equation in this formulation of projectable \Horava  gravity results from
explicit metric variations alone, since $u_a$ is metric-independent, 
and it reads
\begin{equation}
\fom_{ab}-\tilde\lambda u_a u_b=0.
\end{equation}
Using the definition of $\tilde\mu$ in \eqref{lambdaPF} we can rewrite this as
\begin{equation}\label{eqn:projEE}
\fom _{ab}-\frac12\eom^cu_c u_a u_b - \frac12\tilde\mu u_a u_b = 0,
\end{equation}
which looks like the khronometric theory equation \eqref{eqn:foliationEE}
with an additional ``dark matter" component.\footnote{In the formulation
with the integrated constraint, we would find the same Einstein equation,
with the identification of $\tilde\mu = u_c\nabla_d\tilde\lambda^{dc}$} 
We note a peculiar
difference from the 
foliation + clock
theory: $\tilde\mu$ appears
in \eqref{eqn:projEE} with the opposite sign as in \eqref{eqn:clockfoliationEE}.  Thus, in 
projectable \Horava theory, positive $\tilde\mu$ represents \emph{negative} energy 
density, whereas it gives positive energy density in the  
foliation + clock
theory.

The on-shell value of the constraint quantity \eqref{eqn:covconstr} 
associated with the clock field $\psi$ is
\begin{equation}
\constr^{(\psi)}_b 
= \tilde\mu u_b,
\end{equation}
so again we find a single constraint violated, due to the presence of 
the ``dark matter energy density" $\tilde\mu$.

Unlike previous cases considered in this paper, $\tilde\mu$ is generically
not conserved along the aether flow in projectable \Horava gravity.  
The conservation equation comes from 
the clock field equation of motion, which reads
\beq
\nabla_a\left[\eom^a-2\tilde\lambda u^a\right] = 0.
\eeq
Decomposing this equation into parallel and perpendicular components, we find 
evolution equation for $\tilde\mu$,
\beq \label{eqn:projmuconserve}
\lie_u\tilde\mu =  \nabla_a \eom^a_\perp.
\eeq
Since the aether equation of motion is not imposed, 
this means the ``dark matter" may be generated or destroyed  
along the flow of $u^a$.  
Non-conservation of the ``dark matter as an integration constant"
was pointed out in \cite{Mukohyama2009}, 
where it was suggested that this could
provide a mechanism for the generation of dark matter during the early universe. 
In that paper, it was assumed that the theory agrees with general relativity in 
the IR, so that the coupling parameters $c_i$ are zero. 
(Recall that the Lagrange multiplier term results in nonzero $\tilde\mu$
even when the aether couplings are zero.)  In this limit, $\eom^a=0$, 
so we would recover the conservation equation $\lie_u \tilde \mu=0$ were it not
for the higher derivative terms included in the the full \Horava-Lifshitz theory \cite{Mukohyama2009}.
We note that the non-conservation of $\tilde\mu$ is potentially problematic. 
Apparently nothing enforces that
$-\tilde\mu$  remain positive, so instabilities might arise.\footnote{Only when $\tilde\mu$
is conserved due to the aether parameters being zero has it been shown that fluctuations
around positive energy backgrounds have positive energy \cite{Blas2009, Barvinsky2014}.}

\section{Discussion} 

In this paper we studied a variety of aether theories 
including and modifying Einstein-aether theory and
the IR limit of \Horava-Lifshitz gravity (khronometric gravity),
which differ only in how the
aether is constructed from the independently varied fields in the action. When those fields
are scalars, their equations of motion are implied by the other equations of motion, so they
may be regarded as defining fixed background structures. We found that it can be consistent
to include such background structures, and that they often induce extra degrees of 
freedom owing to the loss of diffeomorphism constraints.

The specific structures we considered 
were the 
fixed threading and fixed foliation, as well as 
a fixed clock field that could be included in either the threading
or foliation theories.  We also considered a non-fixed threading, 
described by a vector field rather than a triple of scalar fields.
For the fixed threading theory, the Einstein equation 
was unaltered relative to Einstein-aether theory, 
but the perpendicular aether equation of motion and the corresponding
constraint equations were modified by a constant source term. The foliation theory
without additional structure is equivalent to nonprojectable \Horava gravity.
For this theory, the initial value constraints hold in 
the adapted
gauge, but the Einstein equation differs from 
the threading-type theories since the aether appears naturally as a covector $u_a$. 

The addition of the (fixed) clock field $\psi$ modifies the Einstein equation by an additional
term that has the form of a pressureless dust stress tensor, which can be 
thought of as due to an internal energy density of the aether. We therefore called 
such aethers ``ponderable." The fixed clock
also leads to a violation of the $\psi$ component
of the initial value constraint. \
When a fixed clock is added to the fixed threading theory, we obtain a ``fixed aether"
theory, which is equivalent to describing the aether by a vector field that is not varied in the action.
When a fixed clock is added to the foliation theory and constrained to be constant on the
preferred foliation, the projectable version of \Horava gravity results.

The appearance of a 
dark-matter-like component in the Einstein equation
is also a feature of the recently proposed mimetic dark matter theory \cite{Chamseddine2013}.
In this  
theory, the physical metric $g_{ab}$ is constructed from 
another metric $\bar{g}_{ab}$ and a scalar $\phi$ such that
the gradient $\nabla_a\phi$ is unit by construction, 
$g_{ab} = (\bar{g}^{cd}\nabla_c\phi \nabla_d\phi)\bar{g}_{ab}$.
It was shown in \cite{Golovnev2014} that this theory is equivalent to ordinary Einstein gravity,
supplemented with a scalar field $\phi$ that appears in the action 
only via the constraint term imposing that $\nabla_c\phi$ is unit.  
The discussion of section
\ref{sec:projectable} therefore 
demonstrates that this 
theory is equivalent to projectable \Horava 
gravity with vanishing aether action, that is, with  
the parameters $c_i$ set  
to zero.\footnote{More generally,  all the background clock theories discussed 
here can be formulated using the $\bar g_{ab}$ metric construction 
of  \cite{Chamseddine2013}
(see also \cite{Barvinsky2014} for the case of a vector field),
instead of imposing the unit constraint with a Lagrange multiplier term.
If we define the physical metric
as $g_{ab} = (\bar{g}^{cd}u_c u_d) \bar{g}_{ab}$ for a covariant aether, or 
as $g_{ab} = (\bar{g}_{cd} u^c u^d)^{-1} \bar{g}_{ab}$ for a contravariant aether,
the aether vector automatically has unit norm with respect to the physical metric.
The action depends only on the conformal class of $\bar g_{ab}$, so the 
$\bar g_{ab}$ variation gives a trace free equation, 
$(G^{ab}-T^{ab}) - (G-T) u^a u^b=0$, where $G^{ab}$ and $T^{ab}$ are the variational
derivatives  of the action with respect to $g_{ab}$, and the traces $G$ and $T$ are their
contractions with $g_{ab}$. For the fixed aether theory, there is no $u^a$ variation equation, and instead of the usual Einstein equation we have only this Einstein equation with an additional source with pressureless dust energy-momentum tensor and energy density $G-T$. Thus we have recovered the ponderable aether discussed in the text. 
The equivalence to a mimetic dark matter theory arises in the case when there is no aether action.
Then $T^{ab}$ is just the matter stress tensor and is conserved when the matter satisfies its equation of motion, so the Bianchi identity implies the conservation law $\nabla_a[(G-T) \hat u^a \hat u^b]=0$. The extra term thus behaves in this case as geodesic dark matter dust.  
}
Thus, the mimetic dark matter theory 
can be viewed as  
a special case of the aethereal
theories described here.

The threading theory formalism discussed here
resembles the Lagrangian description of perfect fluids.  
The aether,  constructed
from the comoving potentials $\vphi^I$, acts as a zero temperature ``vacuum fluid", 
according to the thermodynamic relations developed in \cite{Dubovsky2012}.  
The vanishing of the temperature
is closely tied with the enhanced symmetry of the aether fluid, 
which 
includes
all $\vphi^I$-diffeomorphisms rather than only 
the volume
preserving ones. 
It was mentioned in
\cite{Dubovsky2006} that such an enhanced symmetry is not possible
without adding more fields, 
but this is true only for an action that is first order in derivatives.\footnote{With 
the addition of another scalar field it is possible. An example is provided by 
a Lagrangian $F(y)$ that is a function only of the chemical potential $y=u^a \nabla_a\psi$,
where $u^a$ is the fluid velocity \eqref{u(At)}.  This would not contain higher derivative 
terms, and it is invariant under full $\vphi^I$ diffeomorphisms (not just 
volume preserving ones). It also has the chemical shift symmetry, so its symmetries
are the same as those of the fixed aether theory. It can be shown that 
an $F(y)$ Lagrangian possesses the same dynamics as an uncharged perfect fluid.
It cannot produce an equation of state $p=0$, whereas the
formulation using only $\vphi^I$ cannot produce $\rho=0$.}
The lowest order terms in the aether action involve second derivatives of $\vphi^I$.  
A derivative expansion for a fluid action was discussed in \cite{Dubovsky2012,Bhattacharya2013},
and the aethereal terms invariant under all $\vphi^I$-diffeomorphisms appear in the latter reference.

The clock field $\psi$ in the aether theories is
analogous to the phase field introduced in \cite{Dubovsky2012} for fluids
carrying a conserved particle number.  In particular, it possesses the same ``chemical shift'' symmetry, 
$\psi\mapsto\psi + \upsilon(\vphi^I)$, which, in the aethereal case, corresponds to 
a freedom to shift the initial value of the clock 
along each thread.  The scalar $y=u^a\nabla_a\psi$ 
has a thermodynamic interpretation as the chemical potential for 
particles charged under the shift symmetry. 
In our aethereal setting, however, $y$ is fixed everywhere equal to
unity by construction of $u^a$ [see, e.g.\  Eq.\ \eqref{eqn:threadparam}],
and $\psi$ measures proper time along the flow.

Having elucidated the basic structure of these theories, 
it is interesting
 to consider the phenomenological consequences of the 
 presence of the background structures.
In a companion paper \cite{Jacobson2015}, we considered some of the 
astrophysical and cosmological implications of the source densities for the
threading-type theories.  In particular, we examined whether the new component
in the Einstein equation for ponderable aethers with fixed clocks could
play the role of dark matter.  
The two main results of that analysis are that (i) the ``aethereal dark matter fluid" has pressure,
hence does not
seed structure formation on sub-horizon scales, so another dark matter component 
must be present, and (ii) during matter domination, the presence of 
a homogeneous ponderable aether energy density causes problematic
growth of the isocurvature modes on super-horizon scales.  In particular, 
for isocurvature amplitudes of order $10^{-5}$ at radiation-matter equality
(which would be the value expected from inflation \cite{Armendariz-Picon2010}), 
the growth at large scales becomes inconsistent with 
CMB and large scale structure observations 
when the ponderable aether contributes more than $1\%$ of the
homogeneous energy density. 
On the other hand, these results do not apply to the ``dark
matter as an integration constant'' \cite{Mukohyama2009}
in projectable \Horava gravity.  That theory
results from taking the limit $c_a\rightarrow\infty, c_\omega\rightarrow\infty$ of 
Einstein-aether theory \cite{Blas2010, Jacobson2013}.  In that limit the 
dark matter fluid is pressureless, and the 
large-scale isocurvature modes are decaying. 

Finally, it should perhaps 
be emphasized that, in a theory with conserved ``aethereal dark matter" current,
the primordial value of the internal energy density of the aether would 
be driven very nearly to zero if there is an early period of inflation.
This leads to the curious conclusion
that the nondynamical and dynamical
aether theories  
could appear to be
essentially equivalent in their phenomenological predictions.

\begin{acknowledgments}
We are grateful to A. Vikman for helpful correspondence, and to an anonymous
referee for helpful suggestions and comments. 
This research was supported in part by the 
National Science Foundation under grants No. PHY-0903572, PHY-1407744, and PHY11-25915,
and by Perimeter Institute for Theoretical Physics.  Research at 
Perimeter Institute is supported by 
the Government of Canada through Industry Canada and 
by the Province of Ontario through the 
Ministry of Research \& Innovation. 
AS thanks the UCSB Department of Physics for hospitality while this work was being completed.

\end{acknowledgments}

\bibliography{timefiber}

\begin{thebibliography}{33}%
\makeatletter
\providecommand \@ifxundefined [1]{%
 \@ifx{#1\undefined}
}%
\providecommand \@ifnum [1]{%
 \ifnum #1\expandafter \@firstoftwo
 \else \expandafter \@secondoftwo
 \fi
}%
\providecommand \@ifx [1]{%
 \ifx #1\expandafter \@firstoftwo
 \else \expandafter \@secondoftwo
 \fi
}%
\providecommand \natexlab [1]{#1}%
\providecommand \enquote  [1]{``#1''}%
\providecommand \bibnamefont  [1]{#1}%
\providecommand \bibfnamefont [1]{#1}%
\providecommand \citenamefont [1]{#1}%
\providecommand \href@noop [0]{\@secondoftwo}%
\providecommand \href [0]{\begingroup \@sanitize@url \@href}%
\providecommand \@href[1]{\@@startlink{#1}\@@href}%
\providecommand \@@href[1]{\endgroup#1\@@endlink}%
\providecommand \@sanitize@url [0]{\catcode `\\12\catcode `\$12\catcode
  `\&12\catcode `\#12\catcode `\^12\catcode `\_12\catcode `\%12\relax}%
\providecommand \@@startlink[1]{}%
\providecommand \@@endlink[0]{}%
\providecommand \url  [0]{\begingroup\@sanitize@url \@url }%
\providecommand \@url [1]{\endgroup\@href {#1}{\urlprefix }}%
\providecommand \urlprefix  [0]{URL }%
\providecommand \Eprint [0]{\href }%
\providecommand \doibase [0]{http://dx.doi.org/}%
\providecommand \selectlanguage [0]{\@gobble}%
\providecommand \bibinfo  [0]{\@secondoftwo}%
\providecommand \bibfield  [0]{\@secondoftwo}%
\providecommand \translation [1]{[#1]}%
\providecommand \BibitemOpen [0]{}%
\providecommand \bibitemStop [0]{}%
\providecommand \bibitemNoStop [0]{.\EOS\space}%
\providecommand \EOS [0]{\spacefactor3000\relax}%
\providecommand \BibitemShut  [1]{\csname bibitem#1\endcsname}%
\let\auto@bib@innerbib\@empty
\bibitem [{\citenamefont {Boersma}\ and\ \citenamefont
  {Dray}(1995{\natexlab{a}})}]{Boersma1995}%
  \BibitemOpen
  \bibfield  {author} {\bibinfo {author} {\bibfnamefont {S.}~\bibnamefont
  {Boersma}}\ and\ \bibinfo {author} {\bibfnamefont {T.}~\bibnamefont {Dray}},\
  }\bibfield  {title} {\enquote {\bibinfo {title} {{Slicing, threading and
  parametric manifolds}},}\ }\href {\doibase 10.1007/BF02109128} {\bibfield
  {journal} {\bibinfo  {journal} {Gen. Relativ. Gravit.}\ }\textbf {\bibinfo
  {volume} {27}},\ \bibinfo {pages} {319} (\bibinfo {year}
  {1995}{\natexlab{a}})},\ \Eprint {http://arxiv.org/abs/9407020}
  {arXiv:9407020 [gr-qc]} \BibitemShut {NoStop}%
\bibitem [{\citenamefont {Boersma}\ and\ \citenamefont
  {Dray}(1995{\natexlab{b}})}]{Boersma1995a}%
  \BibitemOpen
  \bibfield  {author} {\bibinfo {author} {\bibfnamefont {S.}~\bibnamefont
  {Boersma}}\ and\ \bibinfo {author} {\bibfnamefont {T.}~\bibnamefont {Dray}},\
  }\bibfield  {title} {\enquote {\bibinfo {title} {{Parametric manifolds. I.
  Extrinsic approach}},}\ }\href {\doibase 10.1063/1.531127} {\bibfield
  {journal} {\bibinfo  {journal} {J. Math. Phys.}\ }\textbf {\bibinfo {volume}
  {36}},\ \bibinfo {pages} {1378} (\bibinfo {year} {1995}{\natexlab{b}})},\
  \Eprint {http://arxiv.org/abs/9407011} {arXiv:9407011 [gr-qc]} \BibitemShut
  {NoStop}%
\bibitem [{\citenamefont {Blas}\ \emph {et~al.}(2009)\citenamefont {Blas},
  \citenamefont {Pujol\`{a}s},\ and\ \citenamefont {Sibiryakov}}]{Blas2009}%
  \BibitemOpen
  \bibfield  {author} {\bibinfo {author} {\bibfnamefont {D.}~\bibnamefont
  {Blas}}, \bibinfo {author} {\bibfnamefont {O.}~\bibnamefont {Pujol\`{a}s}}, \
  and\ \bibinfo {author} {\bibfnamefont {S.}~\bibnamefont {Sibiryakov}},\
  }\bibfield  {title} {\enquote {\bibinfo {title} {{On the extra mode and
  inconsistency of Ho\v{r}ava gravity}},}\ }\href {\doibase
  10.1088/1126-6708/2009/10/029} {\bibfield  {journal} {\bibinfo  {journal} {J.
  High Energy Phys.}\ }\textbf {\bibinfo {volume} {2009}},\ \bibinfo {pages}
  {029} (\bibinfo {year} {2009})},\ \Eprint {http://arxiv.org/abs/0906.3046v2}
  {arXiv:0906.3046v2 [hep-th]} \BibitemShut {NoStop}%
\bibitem [{\citenamefont {Blas}\ \emph {et~al.}(2010)\citenamefont {Blas},
  \citenamefont {Pujol\`{a}s},\ and\ \citenamefont {Sibiryakov}}]{Blas2010}%
  \BibitemOpen
  \bibfield  {author} {\bibinfo {author} {\bibfnamefont {D.}~\bibnamefont
  {Blas}}, \bibinfo {author} {\bibfnamefont {O.}~\bibnamefont {Pujol\`{a}s}}, \
  and\ \bibinfo {author} {\bibfnamefont {S.}~\bibnamefont {Sibiryakov}},\
  }\bibfield  {title} {\enquote {\bibinfo {title} {{Consistent Extension of
  Ho\v{r}ava Gravity}},}\ }\href {\doibase 10.1103/PhysRevLett.104.181302}
  {\bibfield  {journal} {\bibinfo  {journal} {Phys. Rev. Lett.}\ }\textbf
  {\bibinfo {volume} {104}},\ \bibinfo {pages} {181302} (\bibinfo {year}
  {2010})},\ \Eprint {http://arxiv.org/abs/0909.3525v1} {arXiv:0909.3525v1
  [hep-th]} \BibitemShut {NoStop}%
\bibitem [{\citenamefont {Jacobson}(2010)}]{Jacobson2010a}%
  \BibitemOpen
  \bibfield  {author} {\bibinfo {author} {\bibfnamefont {T.}~\bibnamefont
  {Jacobson}},\ }\bibfield  {title} {\enquote {\bibinfo {title} {{Extended
  Ho\v{r}ava gravity and Einstein-aether theory}},}\ }\href {\doibase
  10.1103/PhysRevD.81.101502} {\bibfield  {journal} {\bibinfo  {journal} {Phys.
  Rev. D}\ }\textbf {\bibinfo {volume} {81}},\ \bibinfo {pages} {101502}
  (\bibinfo {year} {2010})},\ \Eprint {http://arxiv.org/abs/1001.4823v3}
  {arXiv:1001.4823v3 [hep-th]} \BibitemShut {NoStop}%
\bibitem [{\citenamefont {Ho\v{r}ava}(2009)}]{Horava2009}%
  \BibitemOpen
  \bibfield  {author} {\bibinfo {author} {\bibfnamefont {P.}~\bibnamefont
  {Ho\v{r}ava}},\ }\bibfield  {title} {\enquote {\bibinfo {title} {{Quantum
  gravity at a Lifshitz point}},}\ }\href {\doibase 10.1103/PhysRevD.79.084008}
  {\bibfield  {journal} {\bibinfo  {journal} {Phys. Rev. D}\ }\textbf {\bibinfo
  {volume} {79}},\ \bibinfo {pages} {084008} (\bibinfo {year} {2009})},\
  \Eprint {http://arxiv.org/abs/0901.3775v2} {arXiv:0901.3775v2 [hep-th]}
  \BibitemShut {NoStop}%
\bibitem [{\citenamefont {Jacobson}\ and\ \citenamefont
  {Mattingly}(2001)}]{Jacobson2001}%
  \BibitemOpen
  \bibfield  {author} {\bibinfo {author} {\bibfnamefont {T.}~\bibnamefont
  {Jacobson}}\ and\ \bibinfo {author} {\bibfnamefont {D.}~\bibnamefont
  {Mattingly}},\ }\bibfield  {title} {\enquote {\bibinfo {title} {{Gravity with
  a dynamical preferred frame}},}\ }\href {\doibase 10.1103/PhysRevD.64.024028}
  {\bibfield  {journal} {\bibinfo  {journal} {Phys. Rev. D}\ }\textbf {\bibinfo
  {volume} {64}},\ \bibinfo {pages} {024028} (\bibinfo {year} {2001})},\
  \Eprint {http://arxiv.org/abs/gr-qc/0007031v4} {arXiv:gr-qc/0007031v4}
  \BibitemShut {NoStop}%
\bibitem [{\citenamefont {Jacobson}(2008)}]{Jacobson2008}%
  \BibitemOpen
  \bibfield  {author} {\bibinfo {author} {\bibfnamefont {T.}~\bibnamefont
  {Jacobson}},\ }\bibfield  {title} {\enquote {\bibinfo {title}
  {{Einstein-aether gravity: a status report}},}\ }in\ \href
  {http://arxiv.org/abs/0801.1547} {\emph {\bibinfo {booktitle} {From Quantum
  to Emergent Gravity: Theory and Phenomenology}}}\ (\bibinfo  {publisher}
  {Proceedings of Science},\ \bibinfo {address} {Trieste},\ \bibinfo {year}
  {2008})\ \Eprint {http://arxiv.org/abs/0801.1547} {arXiv:0801.1547 [gr-qc]}
  \BibitemShut {NoStop}%
\bibitem [{\citenamefont {Jacobson}(2014)}]{Jacobson2013}%
  \BibitemOpen
  \bibfield  {author} {\bibinfo {author} {\bibfnamefont {T.}~\bibnamefont
  {Jacobson}},\ }\bibfield  {title} {\enquote {\bibinfo {title} {{Undoing the
  twist: the Ho\v{r}ava limit of Einstein-aether}},}\ }\href {\doibase
  10.1103/PhysRevD.89.081501} {\bibfield  {journal} {\bibinfo  {journal} {Phys.
  Rev. D}\ }\textbf {\bibinfo {volume} {89}},\ \bibinfo {pages} {081501}
  (\bibinfo {year} {2014})},\ \Eprint {http://arxiv.org/abs/1310.5115}
  {arXiv:1310.5115 [gr-qc]} \BibitemShut {NoStop}%
\bibitem [{\citenamefont {Woodard}(2015)}]{Woodard2015}%
  \BibitemOpen
  \bibfield  {author} {\bibinfo {author} {\bibfnamefont {R.~P.}\ \bibnamefont
  {Woodard}},\ }\href {http://arxiv.org/abs/1506.02210} {\enquote {\bibinfo
  {title} {{The Theorem of Ostrogradsky}},}\ } (\bibinfo {year} {2015}),\
  \Eprint {http://arxiv.org/abs/1506.02210} {arXiv:1506.02210 [hep-th]}
  \BibitemShut {NoStop}%
\bibitem [{\citenamefont {Pitts}(2006)}]{BrianPitts2006}%
  \BibitemOpen
  \bibfield  {author} {\bibinfo {author} {\bibfnamefont {B.~J.}\ \bibnamefont
  {Pitts}},\ }\bibfield  {title} {\enquote {\bibinfo {title} {{Absolute objects
  and counterexamples: JonesÐGeroch dust, Torretti constant curvature,
  tetrad-spinor, and scalar density}},}\ }\href {\doibase
  10.1016/j.shpsb.2005.11.004} {\bibfield  {journal} {\bibinfo  {journal}
  {Stud. Hist. Philos. M. P.}\ }\textbf {\bibinfo {volume} {37}},\ \bibinfo
  {pages} {347} (\bibinfo {year} {2006})},\ \Eprint
  {http://arxiv.org/abs/gr-qc/0506102v4} {arXiv:gr-qc/0506102v4} \BibitemShut
  {NoStop}%
\bibitem [{\citenamefont {Pooley}(2015)}]{Pooley2015}%
  \BibitemOpen
  \bibfield  {author} {\bibinfo {author} {\bibfnamefont {O.}~\bibnamefont
  {Pooley}},\ }\href {http://arxiv.org/abs/1506.03512} {\enquote {\bibinfo
  {title} {{Background Independence, Diffeomorphism Invariance, and the Meaning
  of Coordinates}},}\ } (\bibinfo {year} {2015}),\ \Eprint
  {http://arxiv.org/abs/1506.03512} {arXiv:1506.03512 [physics.hist-ph]}
  \BibitemShut {NoStop}%
\bibitem [{\citenamefont {Mukohyama}(2009)}]{Mukohyama2009}%
  \BibitemOpen
  \bibfield  {author} {\bibinfo {author} {\bibfnamefont {S.}~\bibnamefont
  {Mukohyama}},\ }\bibfield  {title} {\enquote {\bibinfo {title} {{Dark matter
  as integration constant in Ho\v{r}ava-Lifshitz gravity}},}\ }\href {\doibase
  10.1103/PhysRevD.80.064005} {\bibfield  {journal} {\bibinfo  {journal} {Phys.
  Rev. D}\ }\textbf {\bibinfo {volume} {80}},\ \bibinfo {pages} {064005}
  (\bibinfo {year} {2009})},\ \Eprint {http://arxiv.org/abs/0905.3563v4}
  {arXiv:0905.3563v4 [hep-th]} \BibitemShut {NoStop}%
\bibitem [{\citenamefont {Seifert}\ and\ \citenamefont
  {Wald}(2007)}]{Seifert2007a}%
  \BibitemOpen
  \bibfield  {author} {\bibinfo {author} {\bibfnamefont {M.~D.}\ \bibnamefont
  {Seifert}}\ and\ \bibinfo {author} {\bibfnamefont {R.~M.}\ \bibnamefont
  {Wald}},\ }\bibfield  {title} {\enquote {\bibinfo {title} {{General
  variational principle for spherically symmetric perturbations in
  diffeomorphism covariant theories}},}\ }\href {\doibase
  10.1103/PhysRevD.75.084029} {\bibfield  {journal} {\bibinfo  {journal} {Phys.
  Rev. D}\ }\textbf {\bibinfo {volume} {75}},\ \bibinfo {pages} {084029}
  (\bibinfo {year} {2007})},\ \Eprint {http://arxiv.org/abs/gr-qc/0612121v2}
  {arXiv:gr-qc/0612121v2} \BibitemShut {NoStop}%
\bibitem [{\citenamefont {Jacobson}(2011)}]{Jacobson2011}%
  \BibitemOpen
  \bibfield  {author} {\bibinfo {author} {\bibfnamefont {T.}~\bibnamefont
  {Jacobson}},\ }\bibfield  {title} {\enquote {\bibinfo {title} {{Initial value
  constraints with tensor matter}},}\ }\href {\doibase
  10.1088/0264-9381/28/24/245011} {\bibfield  {journal} {\bibinfo  {journal}
  {Classical Quant. Grav.}\ }\textbf {\bibinfo {volume} {28}},\ \bibinfo
  {pages} {9} (\bibinfo {year} {2011})},\ \Eprint
  {http://arxiv.org/abs/1108.1496} {arXiv:1108.1496 [gr-qc]} \BibitemShut
  {NoStop}%
\bibitem [{\citenamefont {Carter}\ and\ \citenamefont
  {Quintana}(1972)}]{Carter1972}%
  \BibitemOpen
  \bibfield  {author} {\bibinfo {author} {\bibfnamefont {B.}~\bibnamefont
  {Carter}}\ and\ \bibinfo {author} {\bibfnamefont {H.}~\bibnamefont
  {Quintana}},\ }\bibfield  {title} {\enquote {\bibinfo {title} {{Foundations
  of General Relativistic High-Pressure Elasticity Theory}},}\ }\href {\doibase
  10.1098/rspa.1972.0164} {\bibfield  {journal} {\bibinfo  {journal} {P. Roy.
  Soc. Lon. A Mat.}\ }\textbf {\bibinfo {volume} {331}},\ \bibinfo {pages} {57}
  (\bibinfo {year} {1972})}\BibitemShut {NoStop}%
\bibitem [{\citenamefont {Andersson}\ and\ \citenamefont
  {Comer}(2007)}]{Andersson2007}%
  \BibitemOpen
  \bibfield  {author} {\bibinfo {author} {\bibfnamefont {N.}~\bibnamefont
  {Andersson}}\ and\ \bibinfo {author} {\bibfnamefont {G.~L.}\ \bibnamefont
  {Comer}},\ }\bibfield  {title} {\enquote {\bibinfo {title} {{Relativistic
  Fluid Dynamics: Physics for Many Different Scales}},}\ }\href {\doibase
  10.12942/lrr-2007-1} {\bibfield  {journal} {\bibinfo  {journal} {Living Rev.
  Relativ.}\ }\textbf {\bibinfo {volume} {10}} (\bibinfo {year} {2007}),\
  10.12942/lrr-2007-1}\BibitemShut {NoStop}%
\bibitem [{\citenamefont {Dubovsky}\ \emph {et~al.}(2012)\citenamefont
  {Dubovsky}, \citenamefont {Hui}, \citenamefont {Nicolis},\ and\ \citenamefont
  {Son}}]{Dubovsky2012}%
  \BibitemOpen
  \bibfield  {author} {\bibinfo {author} {\bibfnamefont {S.}~\bibnamefont
  {Dubovsky}}, \bibinfo {author} {\bibfnamefont {L.}~\bibnamefont {Hui}},
  \bibinfo {author} {\bibfnamefont {A.}~\bibnamefont {Nicolis}}, \ and\
  \bibinfo {author} {\bibfnamefont {D.~T.}\ \bibnamefont {Son}},\ }\bibfield
  {title} {\enquote {\bibinfo {title} {{Effective field theory for
  hydrodynamics: Thermodynamics, and the derivative expansion}},}\ }\href
  {\doibase 10.1103/PhysRevD.85.085029} {\bibfield  {journal} {\bibinfo
  {journal} {Phys. Rev. D}\ }\textbf {\bibinfo {volume} {85}},\ \bibinfo
  {pages} {085029} (\bibinfo {year} {2012})},\ \Eprint
  {http://arxiv.org/abs/1107.0731v1} {arXiv:1107.0731v1 [hep-th]} \BibitemShut
  {NoStop}%
\bibitem [{\citenamefont {Speranza}(2015)}]{Jacobson2015}%
  \BibitemOpen
  \bibfield  {author} {\bibinfo {author} {\bibfnamefont {A.~J.}\ \bibnamefont
  {Speranza}},\ }\href {http://arxiv.org/abs/1504.03305} {\enquote {\bibinfo
  {title} {Ponderable aether},}\ } (\bibinfo {year} {2015}),\ \Eprint
  {http://arxiv.org/abs/1504.03305} {arXiv:1504.03305 [gr-qc]} \BibitemShut
  {NoStop}%
\bibitem [{\citenamefont {Chamseddine}\ and\ \citenamefont
  {Mukhanov}(2013)}]{Chamseddine2013}%
  \BibitemOpen
  \bibfield  {author} {\bibinfo {author} {\bibfnamefont {A.~H.}\ \bibnamefont
  {Chamseddine}}\ and\ \bibinfo {author} {\bibfnamefont {V.}~\bibnamefont
  {Mukhanov}},\ }\bibfield  {title} {\enquote {\bibinfo {title} {{Mimetic dark
  matter}},}\ }\href {\doibase 10.1007/JHEP11(2013)135} {\bibfield  {journal}
  {\bibinfo  {journal} {J. High Energy Phys.}\ }\textbf {\bibinfo {volume}
  {2013}},\ \bibinfo {pages} {135} (\bibinfo {year} {2013})},\ \Eprint
  {http://arxiv.org/abs/1308.5410v1} {arXiv:1308.5410v1 [astro-ph.CO]}
  \BibitemShut {NoStop}%
\bibitem [{\citenamefont {Mirzagholi}\ and\ \citenamefont
  {Vikman}(2014)}]{Mirzagholi2014}%
  \BibitemOpen
  \bibfield  {author} {\bibinfo {author} {\bibfnamefont {L.}~\bibnamefont
  {Mirzagholi}}\ and\ \bibinfo {author} {\bibfnamefont {A.}~\bibnamefont
  {Vikman}},\ }\href {http://arxiv.org/abs/1412.7136} {\enquote {\bibinfo
  {title} {{Imperfect Dark Matter}},}\ } (\bibinfo {year} {2014}),\ \Eprint
  {http://arxiv.org/abs/1412.7136} {arXiv:1412.7136 [gr-qc]} \BibitemShut
  {NoStop}%
\bibitem [{\citenamefont {Ho\v{r}ava}\ and\ \citenamefont
  {Melby-Thompson}(2010)}]{Horava2010}%
  \BibitemOpen
  \bibfield  {author} {\bibinfo {author} {\bibfnamefont {P.}~\bibnamefont
  {Ho\v{r}ava}}\ and\ \bibinfo {author} {\bibfnamefont {C.~M.}\ \bibnamefont
  {Melby-Thompson}},\ }\bibfield  {title} {\enquote {\bibinfo {title} {{General
  covariance in quantum gravity at a Lifshitz point}},}\ }\href {\doibase
  10.1103/PhysRevD.82.064027} {\bibfield  {journal} {\bibinfo  {journal} {Phys.
  Rev. D}\ }\textbf {\bibinfo {volume} {82}},\ \bibinfo {pages} {064027}
  (\bibinfo {year} {2010})},\ \Eprint {http://arxiv.org/abs/1007.2410v2}
  {arXiv:1007.2410v2 [hep-th]} \BibitemShut {NoStop}%
\bibitem [{\citenamefont {Xu}\ and\ \citenamefont {Ho\v{r}ava}(2010)}]{Xu2010}%
  \BibitemOpen
  \bibfield  {author} {\bibinfo {author} {\bibfnamefont {C.}~\bibnamefont
  {Xu}}\ and\ \bibinfo {author} {\bibfnamefont {P.}~\bibnamefont
  {Ho\v{r}ava}},\ }\bibfield  {title} {\enquote {\bibinfo {title} {{Emergent
  gravity at a Lifshitz point from a Bose liquid on the lattice}},}\ }\href
  {\doibase 10.1103/PhysRevD.81.104033} {\bibfield  {journal} {\bibinfo
  {journal} {Phys. Rev. D}\ }\textbf {\bibinfo {volume} {81}},\ \bibinfo
  {pages} {104033} (\bibinfo {year} {2010})},\ \Eprint
  {http://arxiv.org/abs/1003.0009v1} {arXiv:1003.0009v1 [hep-th]} \BibitemShut
  {NoStop}%
\bibitem [{\citenamefont {Anderson}\ \emph {et~al.}(2012)\citenamefont
  {Anderson}, \citenamefont {Carlip}, \citenamefont {Cooperman}, \citenamefont
  {Ho\v{r}ava}, \citenamefont {Kommu},\ and\ \citenamefont
  {Zulkowski}}]{Anderson2012}%
  \BibitemOpen
  \bibfield  {author} {\bibinfo {author} {\bibfnamefont {C.}~\bibnamefont
  {Anderson}}, \bibinfo {author} {\bibfnamefont {S.~J.}\ \bibnamefont
  {Carlip}}, \bibinfo {author} {\bibfnamefont {J.~H.}\ \bibnamefont
  {Cooperman}}, \bibinfo {author} {\bibfnamefont {P.}~\bibnamefont
  {Ho\v{r}ava}}, \bibinfo {author} {\bibfnamefont {R.~K.}\ \bibnamefont
  {Kommu}}, \ and\ \bibinfo {author} {\bibfnamefont {P.~R.}\ \bibnamefont
  {Zulkowski}},\ }\bibfield  {title} {\enquote {\bibinfo {title} {{Quantizing
  Ho\v{r}ava-Lifshitz gravity via causal dynamical triangulations}},}\ }\href
  {\doibase 10.1103/PhysRevD.85.044027} {\bibfield  {journal} {\bibinfo
  {journal} {Phys. Rev. D}\ }\textbf {\bibinfo {volume} {85}},\ \bibinfo
  {pages} {044027} (\bibinfo {year} {2012})},\ \Eprint
  {http://arxiv.org/abs/1111.6634v1} {arXiv:1111.6634v1 [hep-th]} \BibitemShut
  {NoStop}%
\bibitem [{\citenamefont {Jacobson}\ and\ \citenamefont
  {Speranza}(2014)}]{Jacobson2014}%
  \BibitemOpen
  \bibfield  {author} {\bibinfo {author} {\bibfnamefont {T.}~\bibnamefont
  {Jacobson}}\ and\ \bibinfo {author} {\bibfnamefont {A.~J.}\ \bibnamefont
  {Speranza}},\ }\href {http://arxiv.org/abs/1405.6351} {\enquote {\bibinfo
  {title} {{Comment on `Scalar Einstein-Aether theory'}},}\ } (\bibinfo {year}
  {2014}),\ \Eprint {http://arxiv.org/abs/1405.6351v2} {arXiv:1405.6351v2
  [gr-qc]} \BibitemShut {NoStop}%
\bibitem [{\citenamefont {Lim}\ \emph {et~al.}(2010)\citenamefont {Lim},
  \citenamefont {Sawicki},\ and\ \citenamefont {Vikman}}]{Lim2010}%
  \BibitemOpen
  \bibfield  {author} {\bibinfo {author} {\bibfnamefont {E.~A.}\ \bibnamefont
  {Lim}}, \bibinfo {author} {\bibfnamefont {I.}~\bibnamefont {Sawicki}}, \ and\
  \bibinfo {author} {\bibfnamefont {A.}~\bibnamefont {Vikman}},\ }\bibfield
  {title} {\enquote {\bibinfo {title} {{Dust of dark energy}},}\ }\href
  {\doibase 10.1088/1475-7516/2010/05/012} {\bibfield  {journal} {\bibinfo
  {journal} {J. Cosmol. Astropart. P.}\ }\textbf {\bibinfo {volume} {2010}},\
  \bibinfo {pages} {012} (\bibinfo {year} {2010})},\ \Eprint
  {http://arxiv.org/abs/1003.5751v2} {arXiv:1003.5751v2 [astro-ph.CO]}
  \BibitemShut {NoStop}%
\bibitem [{\citenamefont {Haghani}\ \emph {et~al.}(2014)\citenamefont
  {Haghani}, \citenamefont {Harko}, \citenamefont {Sepangi},\ and\
  \citenamefont {Shahidi}}]{Haghani2014a}%
  \BibitemOpen
  \bibfield  {author} {\bibinfo {author} {\bibfnamefont {Z.}~\bibnamefont
  {Haghani}}, \bibinfo {author} {\bibfnamefont {T.}~\bibnamefont {Harko}},
  \bibinfo {author} {\bibfnamefont {H.~R.}\ \bibnamefont {Sepangi}}, \ and\
  \bibinfo {author} {\bibfnamefont {S.}~\bibnamefont {Shahidi}},\ }\href
  {http://arxiv.org/abs/1404.7689} {\enquote {\bibinfo {title} {{The Scalar
  Einstein-Aether theory}},}\ } (\bibinfo {year} {2014}),\ \Eprint
  {http://arxiv.org/abs/1404.7689v3} {arXiv:1404.7689v3 [gr-qc]} \BibitemShut
  {NoStop}%
\bibitem [{\citenamefont {Capela}\ and\ \citenamefont
  {Ramazanov}(2014)}]{Capela2014}%
  \BibitemOpen
  \bibfield  {author} {\bibinfo {author} {\bibfnamefont {F.}~\bibnamefont
  {Capela}}\ and\ \bibinfo {author} {\bibfnamefont {S.}~\bibnamefont
  {Ramazanov}},\ }\href {http://arxiv.org/abs/1412.2051} {\enquote {\bibinfo
  {title} {{Modified Dust and the Small Scale Crisis in CDM}},}\ } (\bibinfo
  {year} {2014}),\ \Eprint {http://arxiv.org/abs/1412.2051v2}
  {arXiv:1412.2051v2 [astro-ph.CO]} \BibitemShut {NoStop}%
\bibitem [{\citenamefont {Barvinsky}(2014)}]{Barvinsky2014}%
  \BibitemOpen
  \bibfield  {author} {\bibinfo {author} {\bibfnamefont {A.}~\bibnamefont
  {Barvinsky}},\ }\bibfield  {title} {\enquote {\bibinfo {title} {{Dark matter
  as a ghost free conformal extension of Einstein theory}},}\ }\href {\doibase
  10.1088/1475-7516/2014/01/014} {\bibfield  {journal} {\bibinfo  {journal} {J.
  Cosmol. Astropart. P.}\ }\textbf {\bibinfo {volume} {2014}},\ \bibinfo
  {pages} {014} (\bibinfo {year} {2014})},\ \Eprint
  {http://arxiv.org/abs/1311.3111v1} {arXiv:1311.3111v1 [hep-th]} \BibitemShut
  {NoStop}%
\bibitem [{\citenamefont {Golovnev}(2014)}]{Golovnev2014}%
  \BibitemOpen
  \bibfield  {author} {\bibinfo {author} {\bibfnamefont {A.}~\bibnamefont
  {Golovnev}},\ }\bibfield  {title} {\enquote {\bibinfo {title} {{On the
  recently proposed mimetic Dark Matter}},}\ }\href {\doibase
  10.1016/j.physletb.2013.11.026} {\bibfield  {journal} {\bibinfo  {journal}
  {Phys. Lett. B}\ }\textbf {\bibinfo {volume} {728}},\ \bibinfo {pages} {39}
  (\bibinfo {year} {2014})},\ \Eprint {http://arxiv.org/abs/1310.2790v2}
  {arXiv:1310.2790v2 [gr-qc]} \BibitemShut {NoStop}%
\bibitem [{\citenamefont {Dubovsky}\ \emph {et~al.}(2006)\citenamefont
  {Dubovsky}, \citenamefont {Gr\'{e}goire}, \citenamefont {Nicolis},\ and\
  \citenamefont {Rattazzi}}]{Dubovsky2006}%
  \BibitemOpen
  \bibfield  {author} {\bibinfo {author} {\bibfnamefont {S.}~\bibnamefont
  {Dubovsky}}, \bibinfo {author} {\bibfnamefont {T.}~\bibnamefont
  {Gr\'{e}goire}}, \bibinfo {author} {\bibfnamefont {A.}~\bibnamefont
  {Nicolis}}, \ and\ \bibinfo {author} {\bibfnamefont {R.}~\bibnamefont
  {Rattazzi}},\ }\bibfield  {title} {\enquote {\bibinfo {title} {{Null energy
  condition and superluminal propagation}},}\ }\href {\doibase
  http://dx.doi.org/10.1088/1126-6708/2006/03/025} {\bibfield  {journal}
  {\bibinfo  {journal} {J. High Energy Phys.}\ }\textbf {\bibinfo {volume}
  {2006}},\ \bibinfo {pages} {025} (\bibinfo {year} {2006})},\ \Eprint
  {http://arxiv.org/abs/hep-th/0512260v2} {arXiv:hep-th/0512260v2} \BibitemShut
  {NoStop}%
\bibitem [{\citenamefont {Bhattacharya}\ \emph {et~al.}(2013)\citenamefont
  {Bhattacharya}, \citenamefont {Bhattacharyya},\ and\ \citenamefont
  {Rangamani}}]{Bhattacharya2013}%
  \BibitemOpen
  \bibfield  {author} {\bibinfo {author} {\bibfnamefont {J.}~\bibnamefont
  {Bhattacharya}}, \bibinfo {author} {\bibfnamefont {S.}~\bibnamefont
  {Bhattacharyya}}, \ and\ \bibinfo {author} {\bibfnamefont {M.}~\bibnamefont
  {Rangamani}},\ }\bibfield  {title} {\enquote {\bibinfo {title}
  {{Non-dissipative hydrodynamics: effective actions versus entropy
  current}},}\ }\href {\doibase 10.1007/JHEP02(2013)153} {\bibfield  {journal}
  {\bibinfo  {journal} {J. High Energy Phys.}\ }\textbf {\bibinfo {volume}
  {2013}},\ \bibinfo {pages} {153} (\bibinfo {year} {2013})},\ \Eprint
  {http://arxiv.org/abs/1211.1020v3} {arXiv:1211.1020v3 [hep-th]} \BibitemShut
  {NoStop}%
\bibitem [{\citenamefont {Armendariz-Picon}\ \emph {et~al.}(2010)\citenamefont
  {Armendariz-Picon}, \citenamefont {Sierra},\ and\ \citenamefont
  {Garriga}}]{Armendariz-Picon2010}%
  \BibitemOpen
  \bibfield  {author} {\bibinfo {author} {\bibfnamefont {C.}~\bibnamefont
  {Armendariz-Picon}}, \bibinfo {author} {\bibfnamefont {N.~F.}\ \bibnamefont
  {Sierra}}, \ and\ \bibinfo {author} {\bibfnamefont {J.}~\bibnamefont
  {Garriga}},\ }\bibfield  {title} {\enquote {\bibinfo {title} {{Primordial
  perturbations in Einstein-Aether and BPSH theories}},}\ }\href {\doibase
  10.1088/1475-7516/2010/07/010} {\bibfield  {journal} {\bibinfo  {journal} {J.
  Cosmol. Astropart. P.}\ }\textbf {\bibinfo {volume} {2010}},\ \bibinfo
  {pages} {010} (\bibinfo {year} {2010})},\ \Eprint
  {http://arxiv.org/abs/1003.1283v3} {arXiv:1003.1283v3 [astro-ph.CO]}
  \BibitemShut {NoStop}%
\end{thebibliography}%

\end{document}